\documentclass[10pt,journal,a4paper,final,oneside,twocolumn]{IEEEtran}
\usepackage{caption}
\usepackage{times}
\usepackage{graphicx}
\usepackage{cite}
\usepackage{citesort}
\usepackage{color}
\usepackage{psfrag}
\usepackage{subfigure}
\usepackage{amssymb}
\usepackage{epsfig}
\usepackage{pifont}
\usepackage{amsmath}
\usepackage{array}
\usepackage{multicol}
\usepackage[T1]{fontenc}
\usepackage{comment}
\usepackage{acronym}
\usepackage{algorithm}
\usepackage{algorithmic}
\usepackage{multirow}
\usepackage{epsfig}
\usepackage{epstopdf} 

\graphicspath{{./figures/}}
\usepackage{algorithm}
\usepackage{algorithmic}
\usepackage{soul, color, xcolor}
\usepackage{float}
\usepackage{stfloats}
\usepackage{fancyhdr}
\usepackage{bbding}
\usepackage{setspace}
\usepackage{color}
\usepackage{amsmath}
\usepackage{graphicx}
\usepackage{epsfig}
\usepackage{epstopdf}
\usepackage{subfigure}

\usepackage{txfonts}
\usepackage{bm}
\usepackage{setspace}
\begin{document}
	\title{Multi-Domain Polarization for Enhancing the Physical Layer Security of MIMO Systems}
	\author{\IEEEauthorblockN{
			Luping Xiang, \emph{Member, IEEE},
			Yao~Zeng, \emph{Student~Member, IEEE},
			Jie~Hu, \emph{Senior~Member, IEEE},
			Kun~Yang, \emph{Fellow, IEEE}
			and~Lajos~Hanzo}, \emph{Life Fellow, IEEE}
		\vspace{-0.5 cm}\\
		
		\thanks{This work was supported in part by MOST Major Research and Development Project under Grant 2021YFB2900204; in part by the Sichuan Major R\&D Project under Grant 22QYCX0168; in part by the Sichuan Science and Technology Program under Grant 2022YFH0022 and Grant 2023NSFSC1375; in part by the Natural Science Foundation of China under Grant 62132004, Grant 61971102 and Grant 62301122; in part by the Stable Supporting Fund of National Key Laboratory of Underwater Acoustic Technology; and in part by the Key Research and Development Program of Zhejiang Province under Grant 2022C01093. (\textit{Corresponding Author: Jie Hu.})
			
			Luping Xiang, Yao Zeng and Jie Hu are with the School of Information and Communication Engineering, University of Electronic Science and Technology of China, Chengdu 611731, China, email: luping.xiang@uestc.edu.cn, 202122010522@std.uestc.edu.cn, hujie@uestc.edu.cn. 
					
			Kun Yang is with the School of Computer Science and Electronic Engineering, University of Essex, Essex CO4 3SQ, U.K., e-mail: kunyang@essex.ac.uk.
			
			Lajos Hanzo is with the School of Electronics and Computer Science, University of
			Southampton, Southampton SO171BJ, U.K., e-mail: lh@ecs.soton.ac.uk
		}
	}
	\maketitle
	
	\begin{abstract}
		 A novel Physical Layer Security (PLS) framework is conceived for enhancing the security of the wireless communication systems by exploiting multi-domain polarization in Multiple-Input Multiple-Output (MIMO) systems. We design a sophisticated key generation scheme based on multi-domain polarization, and the corresponding receivers. An in-depth analysis of the system's secrecy rate is provided, demonstrating the confidentiality of our approach in the presence of eavesdroppers having strong computational capabilities. More explicitly, our simulation results and theoretical analysis corroborate the advantages of the proposed scheme in terms of its bit error rate (BER), block error rate (BLER), and maximum achievable secrecy rate. Our findings indicate that the innovative PLS framework effectively enhances the security and reliability of wireless communication systems. For instance, in a $4\times4$ MIMO setup, the proposed PLS strategy exhibits an improvement of $2$dB compared to conventional MIMO, systems at a BLER of $2\cdot 10^{-5}$ while the eavesdropper's BLER reaches $1$.
	\end{abstract}
	\section{Introduction}
	
	To enhance the security of wireless communication systems, traditional approaches have primarily relied on secret key based encryption techniques at the network layer. However, the high computational burden of these methods has prompted researchers to explore secure transmission methods at the physical layer (PHY) \cite{ref1,ref2}. Physical layer security (PLS) based mechanisms can be broadly categorized into two groups: keyless PLS transmission techniques based on Wyner's theory \cite{ref3} and key-based PLS transmission techniques rooted in Maurer's theory \cite{Maurer1}. By appropriately integrating these techniques with modulation schemes and channel coding, the security of the system can be improved, while maintaining communication efficiency.
	
	Keyless PLS techniques by definition operate without the need for a key, utilizing sophisticated signal processing methods to degrade the eavesdropper's (E) channel state, while simultaneously enhancing the quality of the legitimate communication channel. The concept of constructive interference, introduced in \cite{ref5}, relies on the transmission of directional artificial noise (AN) to interfere with E. In \cite{ref6}, symbol-level transmit pre-encoders (TPC) are employed for reducing the transmitter's energy consumption and for enhancing the system's overall performance while jamming E. Considering angular errors, Hu \textit{et al.} \cite{ref7} derive a closed-form expression for the AN projection matrix, assuming realistic directional angular estimation errors obeying a uniform distribution within a practical range. Xu \textit{et al.} \cite{AN-Based-VS} designs an effective Artificial Noise Assisted Security Scheme (ANAS), relying on two phases of transmission: in Phase $1$, the legitimate parties send two independent artificial noise sequences (ANs), while in Phase $2$, the transmitter superimposes the ANs received in Phase $1$ on the signals and transmits the resultant sequences mixed signal. Secure communication is achieved since the ANs superimposed on the legitimator, signal in phase $2$ can be effectively cancelled by the legitimate receiver while still interfering with the eavesdropper. Shu \textit{et al.} \cite{ref8} present a robust, AN-based multi-beam broadcast system capable of improving both the security and the rate. Although AN-based keyless designs succeed in increasing the secure transmission rates, this is achieved at the cost of increased complexity and peak to average power ratio (PAPR).
	
	The family of key-based PLS transmission techniques has also garnered interest from numerous researchers \cite{ref9,ref10}. Key generation methods exploit the random physical layer attributes of the channel \cite{ref11} to prevent E from gleaning confidential information from the legitimate links \cite{ref12,ref13,CSI-Based-VS}. The legitimate user employs traditional channel estimation techniques for acquiring the channel state information (CSI) of the legitimate link and subsequently generates the physical layer key \cite{ref14,ref15}. By contrast, E is unable to access the CSI of the legitimate link and the associated key. However, CSI-based key generation schemes are challenging to implement in practice due to biases introduced by channel estimation. This issue has been mitigated through the development of high-performance secure channel coding techniques \cite{ref16}.
		
\begin{table*}[!htb]
		\centering
		\caption{Boldly contrasting our novelty to the literature}
		\label{tab1 contributions}
		\resizebox{\linewidth}{!}{
				\begin{tabular}{lccccccccccc}
					\hline
					Contributions                                       & ours &\cite{ref1,ref2} & \cite{ref3} & \cite{ref4,Maurer1,Maurer2,Maurer3} & \cite{ref5,ref6} & \cite{ref7,ref8} & \cite{ref11} & \cite{ref14,ref15} & \cite{ref20} & \cite{ref23} & \cite{ref25} \\ \hline
					Multiple mapping patterns                           & \CheckmarkBold    &  & & &  &      &      &  \Checkmark    &       &        &        \\ \hline
					Physical layer security (PLS)                        & \CheckmarkBold  & \Checkmark & \Checkmark & \Checkmark &\Checkmark  & \Checkmark    & \Checkmark   &\Checkmark    & \Checkmark      & \Checkmark     & \Checkmark        \\ \hline
					Reduce receiver latency                             & \CheckmarkBold   & & & &    & \Checkmark    &    &      &      &       & \Checkmark        \\ \hline
					Secrecy rate analysis                               & \CheckmarkBold  & & & \Checkmark   &    & \Checkmark    & \Checkmark    & \Checkmark   & \Checkmark      & \Checkmark     & \Checkmark
					\\ \hline
						MIMO polarization                                   & \CheckmarkBold  & & &  &    &      &      &      &        &        &          \\ \hline
						Detection of sequential mapping coding construction & \CheckmarkBold    &  & & &  &      &      &      &        &        &         \\ \hline
			\end{tabular}}
		
\end{table*}

	In conventional communication systems, coding and encryption are treated as separate processes, where physical layer coding is harnessed for enhancing the reliability \cite{ref17}, while upper layer encryption is used for ensuring security \cite{ref18}. For circumventing the weaknesses of upper layer encryption, researchers have embarked on investigating the joint design of coding and encryption at the physical layer \cite{ref19}. This approach is eminently suitable for wireless channels upon using appropriate coding schemes, for simultaneously improving the legitimate link and for preventing E from accessing any confidential information. Powerful low-density parity-check (LDPC) codes are particularly suitable for secure channel coding design. In this context, Li \textit{et al.} \cite{ref20} proposes an LDPC-based McEliece secrecy coding scheme for enhancing the information reliability of legitimate users and the information security against E. Motamedi \textit{et al.} \cite{ref21} examine the-perfect-security' physical layer authentication problem of wireless networks using LDPC codes and hash functions, achieving high authentication rates in the presence of an E having high computational power.
	
	Additionally, the integration of polar codes \cite{ref22} and physical layer security has garnered widespread scholarly attention \cite{PCandPLS1,PCandPLS2}. Polar codes, conceived by Arikan \cite{Arikan}, achieve symmetric capacity for binary input memoryless channels (BMCs). In \cite{ref23}, a concatenated coding scheme combining polar codes and fountain codes is proposed by Yang and Zhuang for memoryless erasure binary eavesdropping channel models, while relying on finite code lengths for ensuring security. Hao \textit{et al.} \cite{ref24} discuss a secure transmission scheme employing two-dimensional polar codes designed for block fading eavesdropping channels, in the face of instantaneous secrecy capacity fluctuations. Bao \textit{et al.} \cite{ref25} combine polar codes with artificial noise to derive upper and lower bounds of the symmetric capacity for polarized bit channels, which benefit the legitimate receiver but not the eavesdropper.
	
	The core of polar code construction lies in the so-called channel polarization processing detailed in \cite{egilmez2019development}. As the coding space dimension approaches infinity, all sub-channels become fully polarized. However, under practical finite code lengths, many sub-channels remain partially polarized, hence impacting the system's secrecy rate. To address this issue, we explore the introduction of multi-domain polarization into physical layer security research. Dai \textit{et al.} \cite{ref26}, guided by the concept of generalized polarization, propose a polarization-coded MIMO model that significantly enhances the benefits of polarization. Explicitly, they demonstrate that multi-domain polarization is eminently suitable for PLS-enhancement.
	
	In this context, we jointly design multi-domain polarization and encryption. On one hand, MIMO detection schemes apply different processing methods and detection orders for the individual spatial layers, resulting in varying signal reliability. Based on this, we design a random detection order based multi-domain polarization model that prevents eavesdroppers from inferring with the legitimate link's MIMO detection mode or multi-domain polarization process, leading to extremely high eavesdropper decoding error rates. On the other hand, since the time-division duplex (TDD) systems' channel reciprocity prevents eavesdroppers from obtaining the legitimate link's instantaneous gain, we partition the gain range into multiple contiguous but non-overlapping intervals. Based on this, we design an instantaneous channel gain mapping based polarization scheme for increasing the randomness of the secret key, hence enhancing the overall system performance, as detailed bellow.

		The key innovations of this scheme are boldly contrasted to the state-of-the-art in Table \ref{tab1 contributions}, which are further detailed as follows:
		
		\begin{itemize}
			\item We propose a novel PLS architecture based on a MIMO scheme, modulation, and multi-domain polarization. This scheme integrates the multi-domain polarization structure with the classic binary polarization coding structure for enhancing the overall system's polarization effect, to a benefit, our solution achieves significant performance improvements over conventional MIMO transmissions. Exploiting the randomness of the MIMO detection order as our secret physical layer key, distinct polarization designs are derived based on different detection orders, yielding unique coding constructions. Since E cannot infer the legitimate link's detection order, it also fails to acquire the corresponding coding construction. This approach enhances the legitimate link's decoding performance and simultaneously it degrades the E link's quality, hence improving the security.
			\item We conceive an instantaneous channel gain based mapping and coding structure. To further enhance the PLS, this method partitions the legitimate link's instantaneous gain into multiple contiguous but non-overlapping intervals, each mapping to a distinct coding construction. By employing the Gaussian approximation (GA) algorithm to match the subchannel reliability, which uses the noise variance of the channel as input to select the most reliable bits, the secret key may be obtained without incurring any additional overhead. Even if E has powerful computational capabilities, it fails to perform accurate decoding. Again, partitioning the legitimate link's gain improves the legitimate link's error correction capability, while degrading the decoding capability of E.
			\item To validate the proposed scheme's confidentiality in the presence of eavesdroppers, we analyze the maximum achievable secrecy rate from various perspectives. Our numerical results confirm the scheme's confidentiality. Furthermore, we evaluate the performance of this approach in terms of both its bit error rate (BER) and block error rate (BLER). Our simulation results demonstrate that even in possession of formidable computing power, eavesdroppers cannot correctly decode a complete data frame. For example, within a $4\times4$ MIMO configuration, the proposed PLS approach attains an SNR enhancement of $2$dB in comparison to conventional MIMO, while the eavesdropper's BLER approaches $100$\% and the legitimate user's BLER is as low as $10^{-5}$.
		\end{itemize}
		
		
		The rest of this paper is composed as follows. In Section \ref{Section PLS design for Multi-domain polarisation MIMO system}, we portray the system model and provide a detailed description of the key generation scheme relying on MIMO based multi-domain polarization. Section \ref{Section Receiver Design} presents the receiver models of both the legitimate user and of the eavesdropper. Subsequently, in Section \ref{Section Secrecy rate analysis}, we analyze the system's secrecy rate. Section \ref{Section Simulation result} provides our simulation results and theoretical analysis. Finally, Section \ref{Section Conclusion} concludes of the paper.
		
		As for our notations, random variables and their actual values are represented by uppercase Roman letters and lowercase letters, respectively. Furthermore, $\mathfrak{R}(x)$ and $\mathfrak{S}(x)$ represent the real and imaginary parts of $x$, respectively. The modulus of $x$ is written as $\|x\|=\sqrt{\mathfrak{R}(x)^{2}+\mathfrak{S}(x)^{2}}$. The calligraphic characters $\mathcal{X}$ and $\mathcal{Y}$ are used to denote sets, and $|\mathcal{X}|$ denotes the number of elements in $\mathcal{X}$. The notation $P(X)$ represents the probability density function (PDF) of random variables, and the probability density function of $X$ is expressed as $p(X|A)$ under the condition of a given $A$. In addition, $\Gamma(n)$ represents the gamma distribution having $n$ degrees of freedom. Matrices and vectors are represented by bold uppercase and lowercase letters, respectively. In particular, $\mathbf{0}_{N\times 1}$ denotes the ($N \times 1$) zero vector and $\mathbf{I}_{N}$ denotes the ($N \times N$) identity matrix. The transpose and conjugate transpose operators are denoted by $(\cdot)^{\mathrm{\prime }}$ and $(\cdot)^{\dagger}$, respectively. Moreover, the element in the $i$-th row and the $j$-th column of matrix $\mathbf{H}$ is written as $h_{i,j}$, while $\mathbf{x}_{1}^{N}$ represents the vector $(x_{1},x_{2},...,x_{N})^{\prime}$. Finally, we employ the notation $E(\cdot)$ to represent the mean operator, and $\|\cdot\|_{F}$ denotes the two-norm operation.
		
		\section{PLS design for Multi-domain polarisation MIMO system}
		\label{Section PLS design for Multi-domain polarisation MIMO system}
		
		This section elaborates on our PLS framework, which relies on MIMO based multi-domain polarization. 
		
		\subsection{Channel Model}
		
		\begin{figure*}[!htb]
			\centering
			\includegraphics[width=0.8\textwidth]{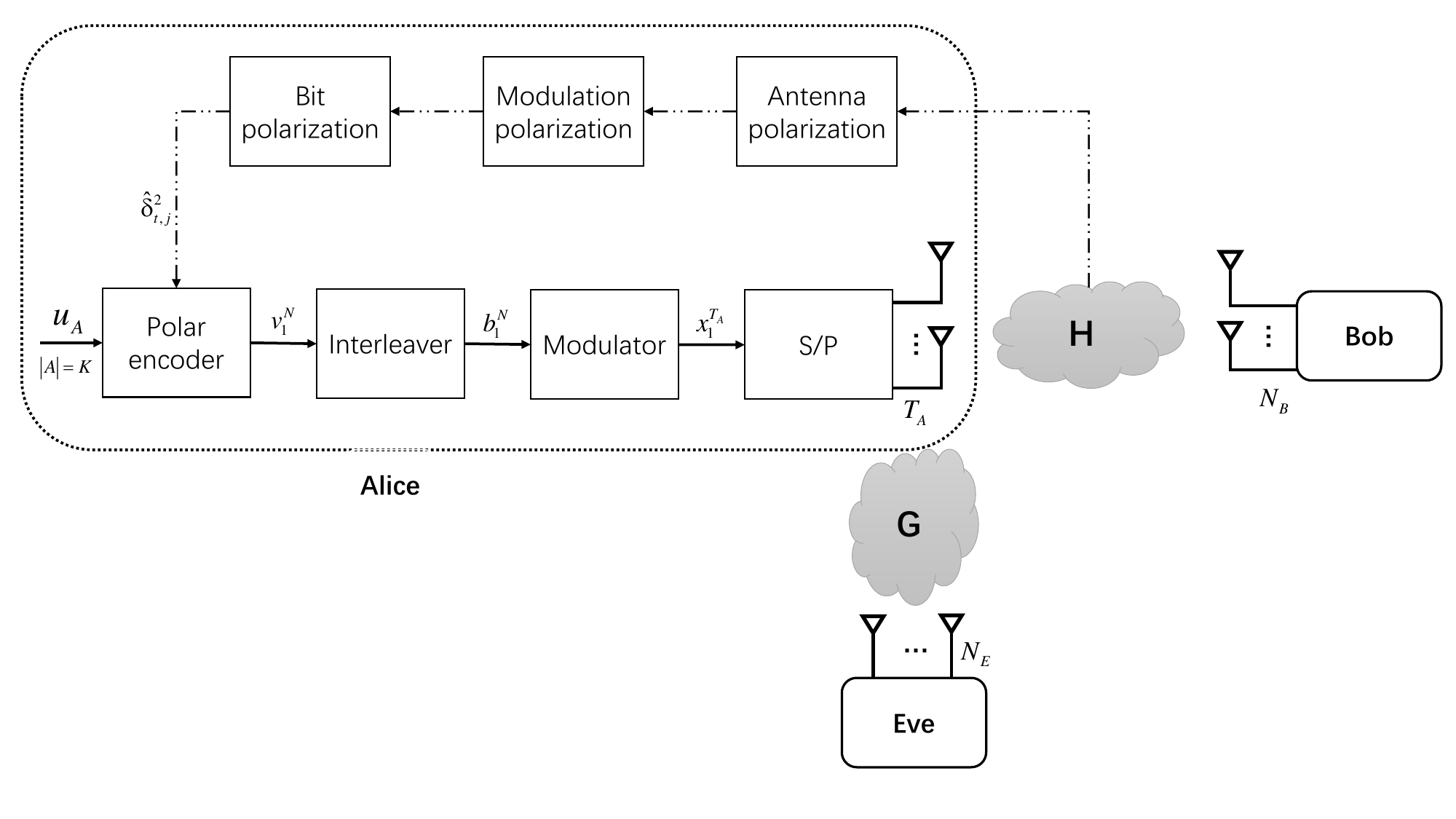}
			\caption{\label{Polar coded MIMO system model}Physical layer security scheme based on MIMO multi-domain polarization.}
		\end{figure*}
		
		Consider the MIMO wiretap channel model depicted in Fig. \ref{Polar coded MIMO system model}. Given a total of $S$ time slots (TS), the transmitter (Alice) sends $K$ information bits to the legitimate user (Bob) after polar coding, interleaving, and modulation using a coding rate of $R=K/N$, where $N$ is the code length. An eavesdropper attempts to intercept the confidential information transmitted via the legitimate link. Alice is equipped with $T_{A}$ transmit antennas (TAs), while Bob and Eve have $N_{B}$ and $N_{E}$ receive antennas (RAs), respectively. The uncorrelated Rayleigh fading channels encountered by the legitimate link and the eavesdropping link are denoted by $\mathbf{H}=\left[\mathbf{h}_{\mathbf{1}}, \mathbf{h}_{\mathbf{2}}, \cdots, \mathbf{h}_{\mathrm{T_{A}}}\right]$ and $\mathbf{G}=\left[\mathbf{g}_{\mathbf{1}}, \mathbf{g}_{\mathbf{2}}, \cdots, \mathbf{g}_{\mathrm{T_{A}}}\right]$, which have sizes of ($N_{B} \times T_{A}$) and ($N_{E} \times T_{A}$), respectively. Each column vector in the matrices $\mathbf{H}$ and $\mathbf{G}$ is expressed as $\mathbf{h}_{\mathrm{t}}=\left[\mathrm{h}_{1 ,\mathrm{t}}, \mathrm{h}_{2 ,\mathrm{t}}, \cdots, \mathrm{h}_{\mathrm{N_{B},t}}\right]^{\prime}$ and $\mathbf{g}_{\mathrm{t}}=\left[\mathrm{g}_{1 ,\mathrm{t}}, \mathrm{g}_{2 ,\mathrm{t}}, \cdots, \mathrm{g}_{\mathrm{N_{E},t}}\right]^{\prime}$, where $t=1,2,...,T_{A}$, respectively. The vectors $\mathbf{h}_{\mathrm{t}}$ and $\mathbf{g}_{\mathrm{t}}$ include the channel coefficients of the link spanning from Alice's $t$-th TA to all RAs of Bob and Eve. Additionally, for any TS, all channel coefficients $\mathrm{h}_{b,t}$ and $\mathrm{g}_{e,t}$ obey $\mathcal{C \mathcal { N }}(0,1)$, where $b$ and $e$ represent the $b$-th row and $e$-th row of $\mathbf{H}$ and $\mathbf{G}$, respectively, while $t$ represents the $t$-th column of $\mathbf{H}$ and $\mathbf{G}$, respectively, with $b=1,2,...,N_{B}, e=1,2,...,N_{E}$.
		
		In a Time Division Duplex (TDD) system, the channel's reciprocity may be exploited without additional resources or overhead, ensuring that Alice and Bob have similar channel coefficients at both end of the link. Therefore, in any TS $s$, the received signal expressions for Bob and Eve are given by:
		
		\begin{equation}
			\label{Y_{B}(s)}
			\mathbf{y}_{1}^{N_{B}}(s)=\mathbf{H}(s)\cdot \mathbf{x}_{1}^{T_{A}}(s)+\mathbf{z}_{1}^{N_{B}}(s),
		\end{equation}
		\begin{equation}
			\label{Y_{E}(s)}
			\mathbf{y}_{1}^{N_{E}}(s)=\mathbf{G}(s) \cdot \mathbf{x}_{1}^{T_{A}}(s)+\mathbf{z}_{1}^{N_{E}}(s).
		\end{equation}
		
		In the $s$-th TS, $s=1,2,...,S$, the vector $\mathbf{y}_{1}^{N{B}}(s)$ of size ($N_{B} \times 1$) represents Bob's received signal, and the vector $\mathbf{y}_{1}^{N{E}}(s)$ of size ($N_{E} \times 1$) contains Eve's received signal. The ($T_{A} \times 1$) vector $\mathbf{x}_{1}^{T{A}}(s)$ represents the symbol transmitted by Alice. Furthermore, the ($N_{B} \times 1$) vector $\mathbf{z}_{1}^{N_{B}}(s)$ and the ($N_{E} \times 1$) vector $\mathbf{z}_{1}^{N_{E}}(s)$ obey the complex Gaussian distributions $\mathcal{C N}\left(\mathbf{0}_{N_{B} \times 1}, \sigma^{2} \mathbf{I}_{N_{B}}\right)$ and $\mathcal{C N}\left(\mathbf{0}_{N_{E} \times 1}, \sigma^{2} \mathbf{I}_{N_{E}}\right)$, containing Bob's and Eve's additive white Gaussian noise (AWGN) components, respectively.
		
		\subsection{Key generation based on multi-domain polarization}
		\label{Key generation based on multi-domain polarization}
		
		\begin{figure*}[!htb]
			\centering
			\label{transmitter}
			\includegraphics[width=0.8\textwidth]{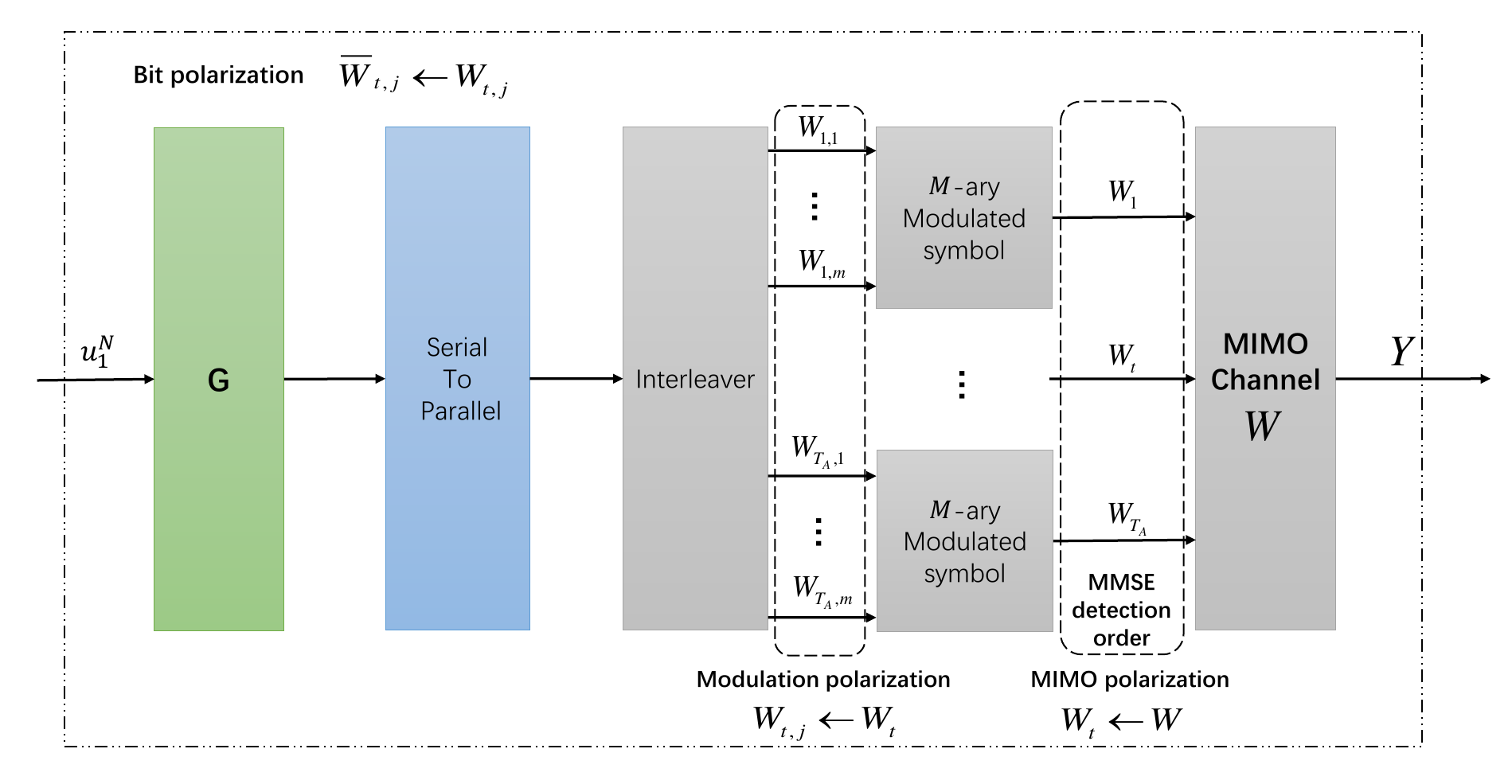}
			\caption{\label{three stage channel transform at Alice}Architecture of MIMO based polarisation at the transmitter.}
		\end{figure*}
		
		Building on the concept of generalized polarization, we aim for enhancing the MIMO transmission efficacy and hence the overall system performance by jointly optimizing the coding and MIMO transmission \cite{ref26}. Again, we propose a MIMO based multi-domain polarization architecture that improves the error correction capability of the legitimate link, while degrading the eavesdropping link's performance. As depicted in Fig. \ref{three stage channel transform at Alice}, the scheme comprises three primary stages \cite{ref26}. In the first stage, MIMO polarization is carried out, which defined as partitioning the original MIMO channel into multiple parallel sub-channels. In the second stage, modulation polarization is carried out following the multi-level coding concept \cite{ref27,ref28} to generate additional bit-based subchannels. Finally, the time slot index is introduced to maximize the system's polarization effect and to select the most reliable bit subchannel for information transmission. Moreover, for avoiding the practical challenges of obtaining the complete legitimate link's CSI, we utilize only the channel's instantaneous gain to design the secure system based on this multi-level polarization approach.
		
		We define the original MIMO channel as $\mathbf{W}: \mathcal{X}^{T_{A}} \mapsto \mathcal{Y}$, where $\mathcal{X}^{T_{A}}$ represents the set of transmitted symbols for each antenna and $|\mathcal{X}^{T_{A}}|=M$, with $M$ being the modulation order, while $\mathcal{Y}$ represents the set of received signals. In TDD systems, the legitimate link's instantaneous channel gain is estimated by the legitimate party. Under such circumstances, the transition probability $\mathbf{W}\left(\mathbf{y}_{1}^{N_{B}}(s) \mid \mathbf{x}_{1}^{T_{A}}(s), \mathbf{H}(s)\right)$ of the legitimate link can be derived according to equation (\ref{Y_{B}(s)}), which can be expressed in the $s$-th TS as \cite{ref26}:
		\begin{align}
			\label{W_B_original}
			\mathbf{W}\left(\mathbf{y}_{1}^{N_{B}}(s) \mid \mathbf{x}_{1}^{T_{A}}(s), \mathbf{H}(s)\right)=\left(\pi \sigma^{2}\right)^{-N_{B}} \cdot \exp \left(-\sum_{i=1}^{N_{B}} \frac{\left\|y_{i}-\tilde{x}_{i}\right\|^{2}}{\sigma^{2}}\right),
		\end{align}
		where $\tilde{x}_{i}$ is the $i$-th element of the ($N_{B} \times 1$) vector $\tilde{\mathbf{x}}_{s}^{N_{B}}(s)=\mathbf{H}(s) \cdot \mathbf{x}_{1}^{T_{A}}(s), i=1,2, \ldots, N_{B}$, $s=1,2,...,S$, while ${y}_{i}$ is the $i$-th element of the ($N_{B} \times 1$) vector $\mathbf{y}_{1}^{N_{B}}(s)$, and $\sigma^{2}$ denotes the noise variance.
		
			\begin{figure*}[!htb]
			\centering
			\subfigure[TA1 polarised versus unpolarised\label{TA1 Polarised}]
			{\includegraphics[width=0.49\textwidth]{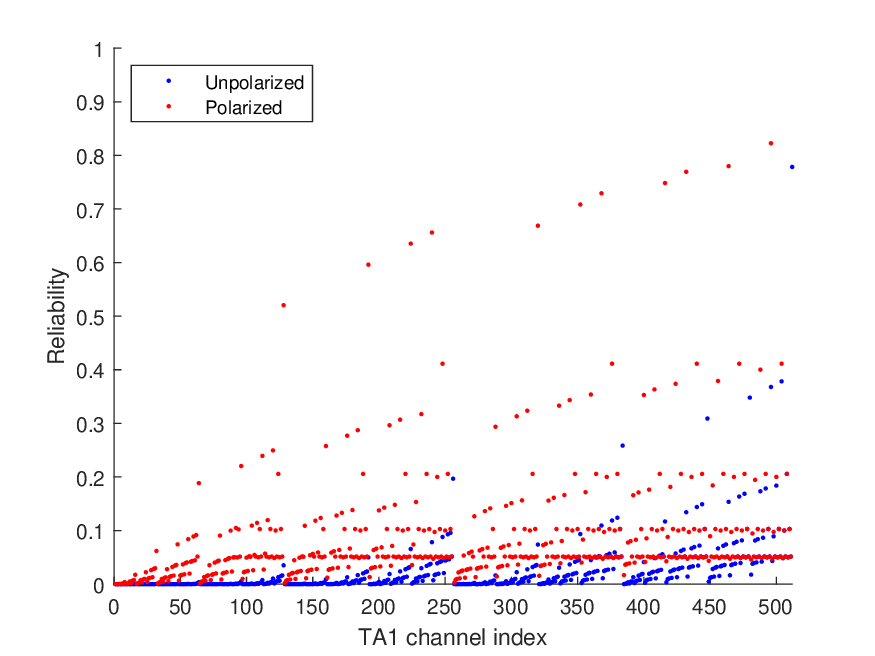}}
			\subfigure[TA2 polarised versus unpolarised\label{TA2 Polarised}]
			{\includegraphics[width=0.49\textwidth]{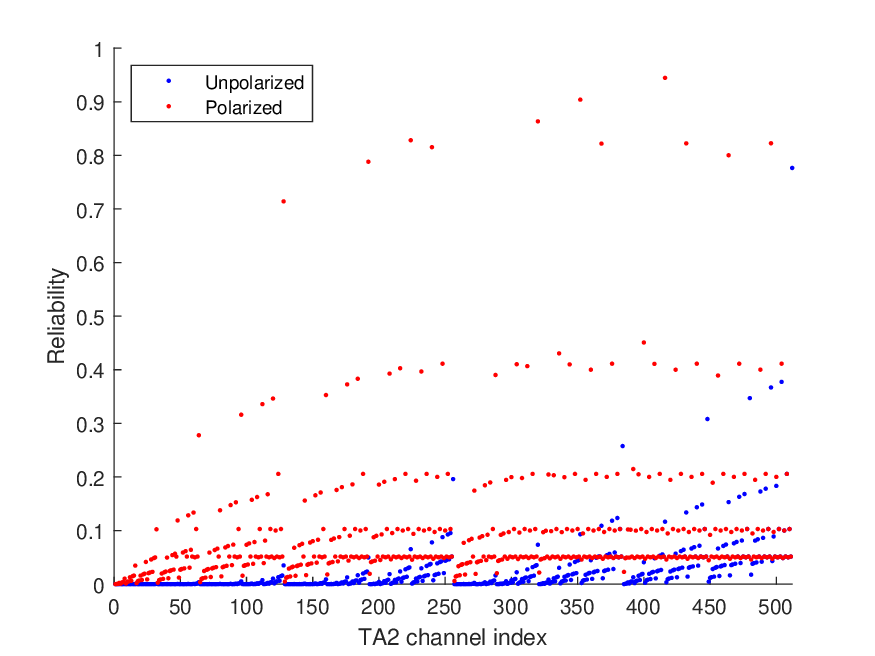}}
			\subfigure[TA3 polarised versus unpolarised\label{TA3 Polarised}]
			{\includegraphics[width=0.49\textwidth]{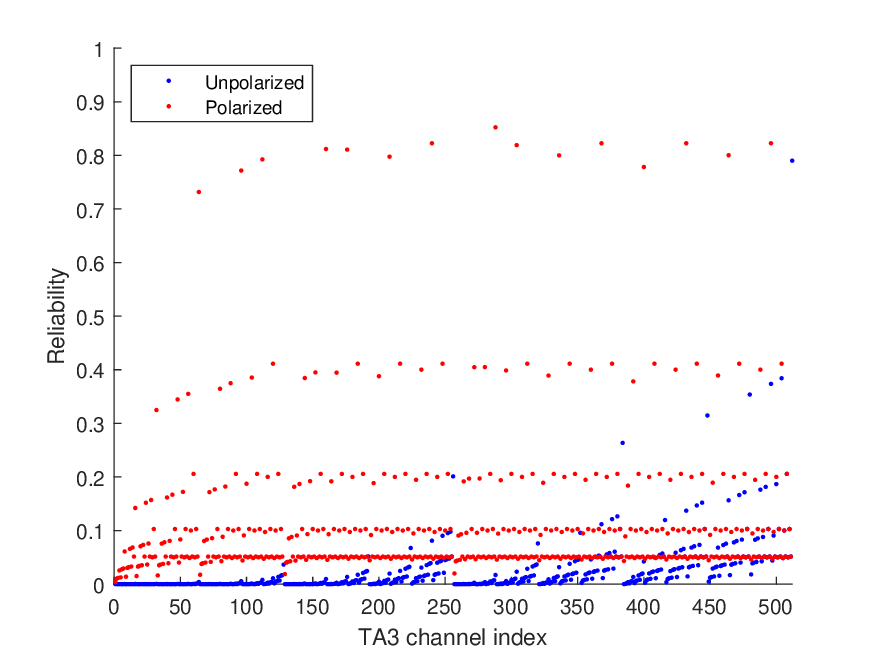}}
			\subfigure[TA4 polarised versus unpolarised\label{TA4 Polarised}]
			{\includegraphics[width=0.49\textwidth]{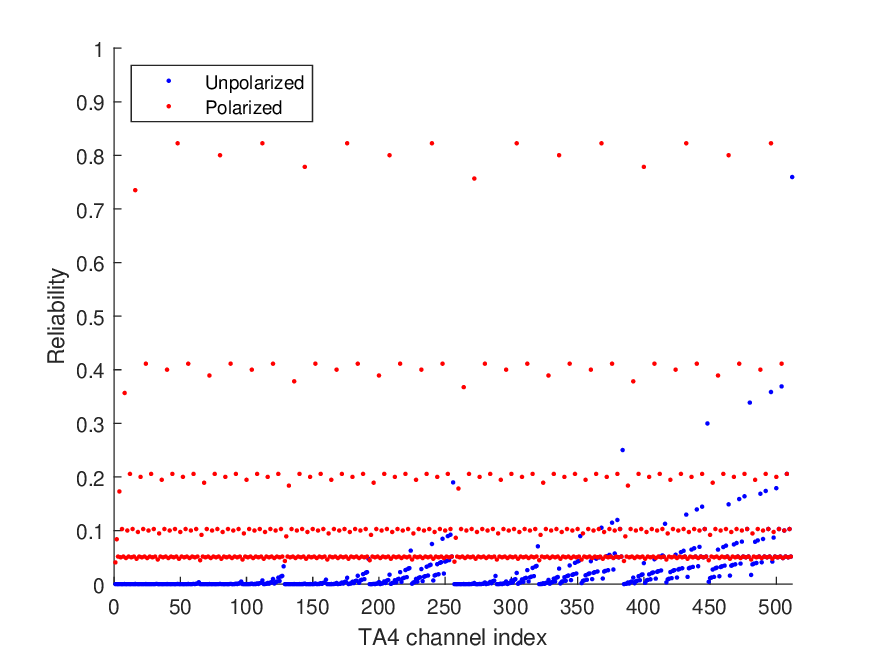}}
			
			\caption{Examples of $4$ $\times$ $4$ MIMO antenna polarisation.}
			\label{TA polorized}
		\end{figure*}
		
		
		At this stage, we perform MIMO polarization. Since the MIMO detection scheme has varying detection orders for each spatial layer, which results in different signal reliability across the individual antennas. For instance, under the linear minimum mean square error (MMSE) successive interference cancellation (SIC) algorithm, the first detected antenna has relatively low reliability due to the interference imposed by the other antennas. Provided that the corresponding symbol was still detected without error, the detected symbol is remodulated and then subtracted both the composite signal, This way the interference is gradually peeled off, thence typically the last detected antenna has the highest reliability due to the absence of interference, which was cancelled by subtracting the remodulated signals of all other RAs. As illustrated in Fig. \ref{TA polorized}, an incremental detection pattern was used in the detection process. In the figure we can see a comparison of the reliability of the different antennas both before and after polarisation. The results show that the average reliability of the antennas after polarisation is significantly higher, further validating the effectiveness of the polarisation technique used. In addition, it should be noted that in the incremental detection mode, the average reliability of the antennas detected in the reverse scan exceeds that of the antennas in the forward scan. This confirms the conclusion of the previous analysis, namely that the interference imposed on the last detected antenna is completely removed. Under this condition, the original MIMO scheme is divided into $T_{A}$ independent sub-channels $\mathbf{W} \rightarrow \mathbf{W}_{t}: \mathcal{X} \mapsto \mathcal{Y}, t=1,2, \ldots, T_{A}$, each associated with different symbol reliability, where $\mathcal{X}$ denotes the set of transmitted symbols. The associated transition probabilities can be further expressed as:
		\begin{equation}
			\centering
			\label{W_B_t}
			\mathbf{W}_{t}\left(\mathbf{y}_{1}^{N_{B}}(s) \mid x_{t}, \mathbf{H}(s)\right)=\sum_{\mathbf{x}_{1}^{T_{A}}(s) \backslash x_{t}} \frac{1}{2^{m(T_{A}-1)}} \cdot \mathbf{W}\left(\mathbf{y}_{1}^{N_{B}}(s) \mid \mathbf{x}_{1}^{T_{A}}(s), \mathbf{H}(s)\right),
		\end{equation}
		where $m=\log_{2}^{M}$ represents the number of bits per $M$-ary quadrature amplitude modulation (QAM) symbol, and $\mathbf{x}_{1}^{T_{A}}(s) \backslash x_{t}$ denotes the subvector of $\mathbf{x}_{1}^{T{A}}(s)$, excluding element $x_{t}$ at the $s$-th TS.
		
		After obtaining $T_{A}$ independent sub-channels having different symbol reliability levels, we proceed to perform modulation polarization \cite{ref28}, introducing polarization effects into the modulated symbol so that each bit sub-channel constituted for example the first or the last bit of the symbol exhibits varying reliability $\mathbf{W} \rightarrow \mathbf{W}_{t}\rightarrow \mathbf{W}_{t,j}:\mathcal{B} \mapsto \mathcal{X} \mapsto \mathcal{Y}, t=1,2, \ldots, T_{A}, j=1,2,...m$, where $\mathcal{B}$ represents the set of transmitted bits $b_{t,j}$. At this point, the transition probability can be written as:
		
		\begin{equation}
			\centering
			\label{W_t_j}
			\begin{aligned}
				&\mathbf{W}_{t, j}\left(\mathbf{y}_{1}^{N_{B}}(s) \mid b_{(t-1) m+j}, \mathbf{H}(s)\right)\\ 
&=\sum_{\mathbf{b}_{(t-1) m+j}^{tm} \backslash b_{(t-1) m+j}}\left(\frac{1}{2^{m-1}} \cdot \mathbf{W}_{t}\left(\mathbf{y}_{1}^{N_{B}}(s) \mid x_{t}, \mathbf{H}(s)\right)\right) \\
				&=\sum_{\mathbf{b}_{(t-1) m+j}^{tm} \backslash b_{(t-1) m+j}, \mathbf{x}_{1}^{T_{A}}(s) \backslash x_{t}}\left(\frac{1}{2^{T_{A} N_{B}-1}} \cdot \mathbf{W}\left(\mathbf{y}_{1}^{N_{B}}(s) \mid \mathbf{x}_{1}^{T_{A}}(s), \mathbf{H}(s)\right)\right),
			\end{aligned}
		\end{equation}
		where $\mathbf{b}_{(t-1) m+j}^{m} \backslash b_{(t-1) m+j}$ represents the bit subvector $\mathbf{b}_{(t-1) m+j}^{m}$ excluding the element $b_{(t-1) m+j}$. Then the binary vector $\mathbf{b}_{(t-1) m+j}^{tm}$ is mapped to the $M$-ary transmitted symbol $x_{t}$ according to the modulation order $M$.
		
		\begin{figure}[!t]
			\centering
			\label{Bit}
			\includegraphics[width=0.49\textwidth]{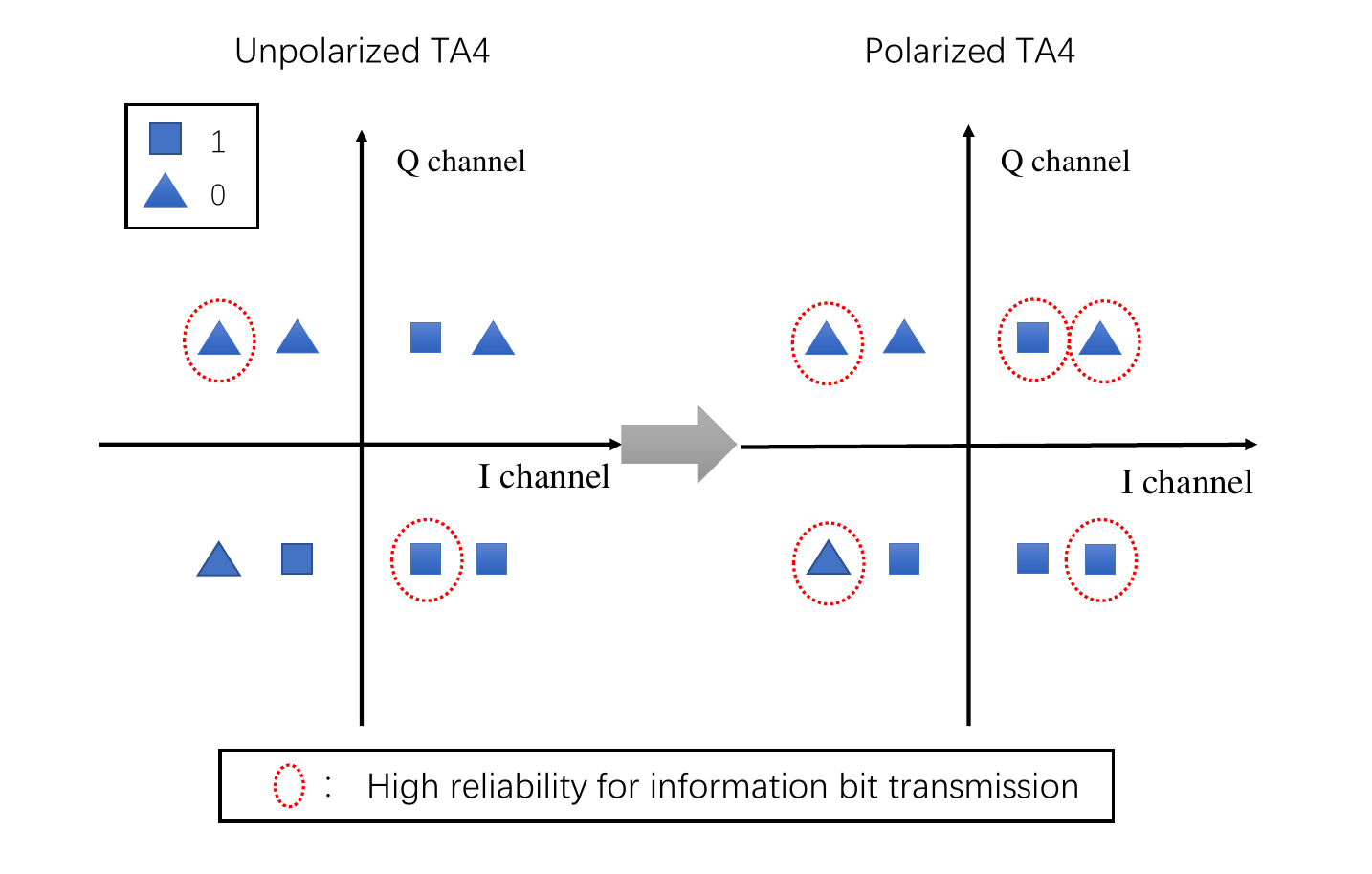}
			\caption{\label{Bit polorized}Examples of bit polarisation of the last antenna detected in increasing order within a $4 \times 4$ MIMO scheme where QPSK is used.}
		\end{figure}
		
		Lastly, we incorporate the time index. Given that the total number of TSs is $S$, the original information sequence is mapped to the corresponding bit sub-channel using polarization coding to state $N$ independent bit sub-channels $\mathbf{W} \rightarrow \mathbf{W}_{t} \rightarrow \mathbf{W}_{t,j} \rightarrow \overline{\mathbf{W}}_{t,j}:\mathcal{U} \mapsto \mathcal{B} \mapsto \mathcal{X} \mapsto \mathcal{Y}$, where $\mathcal{U}$ represents the set of original information bits $u_{t,j}$ having a cardinality of $|\mathcal{U}|=K$. The transition probability can then be expressed as:
		\begin{equation}
			\centering
			\label{W_t_j_time}
			\begin{aligned}
				&\overline{\mathbf{W}}_{t, j}\left(\mathbf{Y}_{B}, \mathbf{u}_{1}^{n-1} \mid u_{n}\right) \\
&=\sum_{\mathbf{u}_{n+1}^{N},\mathbf{b}_{(t-1) m+j}^{tm} \backslash b_{(t-1) m+j}} \frac{\prod_{s=1}^{S} \mathbf{W}_{t,j}\left(\mathbf{y}_{1}^{N_{B}}(s) \mid b_{(t-1) m+j}, \mathbf{H}(s)\right)}{2^{N-1}}.
			\end{aligned}
		\end{equation}
		
		Upon employing the above three-level polarization based channel transformation, the original MIMO channel is polarized into $N$ binary memoryless channels (BMCs). Our MIMO based multi-domain polarization design relies on this cascading principle. The most reliable antenna is selected first through antenna polarization, followed by the selection of the most reliable bit from each RA's modulated symbol. Ultimately, the information bits having the highest reliability are matched across all TSs, resulting in the final polar coding structure. As a benefit of its iterative application \cite{ref29}, the MMSE detection algorithm is used for generating the physical layer key, which is used for mapping the different coding constructs to different antenna detection sequences. In Fig. \ref{Bit polorized}, a toy example is presented to compare the reliability of the antenna that was detected last after polarisation to its unpolarised state, when considering detection executed in ascending order. The figure shows a constellation diagram for QPSK modulation with $8$ points forming $4$ different QPSK symbols. In the unpolarised case, only a limited number of reliable bits can be obtained in the transmitted symbols, the rest being known as frozen bits. However, after polarisation, more reliable bits can be obtained under the same conditions. The reason for this is that after polarisation the average reliability of the bit sub-channel is increased, especially for the symbols transmitted by the last detected antenna, which suffers the least interference. This leads to a significant alteration in the pattern of the polarisation coding structure.
		
		Based on Equations (\ref{Y_{B}(s)}) and (\ref{W_B_t}), the MMSE detector acquires soft estimates of $T_{A}$ independent data streams in the $s$-th TS, after the legitimate party receives the signal associated with the known instantaneous gain of the legitimate link. In this case, the eavesdropper is unable to infer the specific polarization pattern and coding structures since the specific detection method is unattainable. Following the increasing detection order, the soft estimate \cite{ref31} of the $t$-th data stream is formulated as:
		
		\begin{equation}
			\centering
			\label{soft estimation}
			\gamma_{t}(s)=\sum_{\xi=1}^{N_{B}} w_{1, \xi}^{t}(s) \tilde{y}_{\xi}(s),
		\end{equation}
		where $\tilde{y}_{\xi}(s)$ represents the $\xi$-th element of the error vector $\tilde{\mathbf{y}}_{1}^{N_{B}}(s) \triangleq \mathbf{y}_{1}^{N_{B}}(s)-\sum_{\tilde{t}=1}^{t-1} \mathbf{H}_{\tilde{t}}(s) \hat{x}_{\tilde{t}}$ of the received signal in $s$-th time slot. $\mathbf{H}_{\tilde{t}}(s)$ represents a fraction of the original MIMO matrix $\mathbf{H}(s)$ scanning him first column to the $\tilde{t}$-th column, while $\hat{x}_{\tilde{t}}$ represents the symbolic estimate of the $\tilde{t}$-th data stream. Moreover, $w_{1, \xi}^{t}(s)$ represents the $\xi$-th element in the first row of $\mathbf{W}^{t}(s)$, which is the MMSE detection matrix for the $t$-th data stream and its expression is as follows \cite{ref29} :
		
		\begin{equation}
			\centering
			\label{estimation W}
			\mathbf{W}^{t}(s)=\left(\left(\mathbf{H}^{t}(s)\right)^{\dagger} \mathbf{H}^{t}(s)+\sigma^{2} \mathbf{I}_{T_{A}-t+1}\right)^{-1}\left(\mathbf{H}^{t}(s)\right)^{\dagger},
		\end{equation}
		where the matrix $\mathbf{H}^{t}(s)$ represents a fraction of $\mathbf{H}(s)$ scanning him $t$-th column to the $T_{A}$-th column and $\mathbf{I}_{T_{A}-t+1}$ is a unit matrix of size $T_{A}-t+1$.
		
		Considering that the MMSE detection order is random and the transmitter is equipped with $T_{A}$ antennas, the legitimate link will possess $T_{A}!$ distinct detection modes, resulting in $T_{A}!$ unique coding structures for the legitimate link. Under various detection modes, we introduce the equivalent AWGN channel $ \widetilde{\mathbf{W}}_{t, j}$ for transmission. The bit subchannel noise variance, which is obtained under a specific channel fading condition, is transformed into the effective noise variance under the AWGN channel, allowing the same error performance to be achieved under both channels. This implies that the average mutual information (AMI) of the equivalent AWGN channel and the polarized bit subchannel are identical, yielding:
		
		\begin{equation}
			\centering
			\label{equivalent AWGN}
			I\left(\overline{\mathbf{W}}_{t, j}\right)=I\left(\widetilde{\mathbf{W}}_{t, j}\right).
		\end{equation}
		Given the noise variance $\sigma^{2}$, the expression can be written as \cite{ref26}:
		
		\begin{equation}
			\centering
			\label{equivalent variance}
\begin{aligned}
			&I_{\overline{\mathbf{W}}_{t, j}}(\sigma )=I_{\widetilde{\mathbf{W}}_{t, j}}(\sigma_{t,j}) \\
&=\int_{-\infty}^{+\infty} \int_{-\infty}^{+\infty}p(y_{B}) \log _{2}[p(y_{B})] dudv -0.5 \log _{2}\left(2 \pi e\sigma_{t,j}^{2}\right),
\end{aligned}
		\end{equation}
		where $y_{B}$ denotes the signal received by the  legitimate users, and $u=\mathfrak{R}\left(\mathrm{y}_{B}\right), v=\mathfrak{S}\left(\mathrm{y}_{B}\right)$.
		
		In the end, the equivalent noise variance $\sigma_{t,j}^{2}$ of each bit sub-channel is utilized to employ a Gaussian approximation (GA) algorithm for matching the reliability of each sub-channel, as illustrated in Algorithm \ref{algorithm}. Subsequently, confidential information is transmitted with the aid of polarization coding. The distinct detection sequences of the MIMO polarization result in varying antenna reliability levels, leading to different equivalent AWGN variances and coding methods due to the chain reaction of modulation polarization and bit polarization. Again, the random detection order of MIMO polarization determines the secret physical layer key, which is shared by the legitimate link. By contrast, the eavesdropper has only a $1/T_{A}!$ chance of obtaining the correct key. Even if E tentatively tries all possible detection orders, it still cannot determine the correct decoding result. The reason for this is that the detection order determined only ranks the reliability of the antenna and does not give a specific coding structure, which substantially increases the error probability of E. This approach significantly enhances the performance of the legitimate link with the aid of our specific MIMO polarization design, but also considerably degrades the decoding performance of the eavesdropper.
		
		\subsection{Channel gain segmentation design}
		
		\begin{algorithm}[!t]
			\caption{Generation of Coding Construction Scheme $\mathbf{u}_{\mathcal{F}}^{(p)}$}
			\label{algorithm}
			\begin{algorithmic}[1] 
				\REQUIRE ~~\\ 
				Code length $N$\\
				Number of transmitting antennas $T_{A}$\\
				Modulation order $M$\\
				Number of~channel interval $P$\\
				Channel gain $\mu_{t} ,t=1,2,...T_{A}$\\
				Equivalent AWGN noise variance $\delta_{t,j}^{2}$\\
				Length of information bits $K$\\ 
				\ENSURE ~~\\ 
				Frozen bit pattern $\mathbf{u}_{\mathcal{F}}^{(p)}$
				\FOR{$1 \leq t \leq T_{A} $}
				\STATE Calculate $P$ different channel intervals $\left[\varphi_{t,p-1} ,\varphi_{t,p}\right)$ for $t$-th antenna according to~(\ref{P of gain});
				\STATE  Obtain $\left[\phi_{t,p-1},\phi_{t,p}\right)$ by matching the~channel interval with $\mu_{t}$;
				\FOR{$1 \leq j \leq log2^{M}$}
				\STATE Calculate initial $\alpha_{t,j}=\left({\left(\frac{\textit{1}}{\delta_{t,j}^{2}}\right)} \cdot\left(\frac{\phi_{t,p-1}+\phi_{t,p}}{2}\right)\right)^{-\frac{1}{2}}$;
				\STATE Initialize the~LLR mean value of the MIMO channel $\mathbf{W}$ $m_{t,j}^{(1)}=\frac{2}{\alpha_{t,j}^{2}}$;
				\FOR{$0 \leq l \leq n-1 $}
				\STATE Calculate the~mean LLR $m_{{t,2^{n}}}^{(i)}$ of~the~subchannel iteratively according to~\cite{ref29}; 
				\FOR{$1 \leq i\leq 2^{j}$}
				\STATE $m_{t,2^{j+1}}^{(2 i-1)}=\phi^{-1}\left[1-\left(1-\phi\left(m_{t,2^{j}}^{(i)}\right)\right)^{2}\right]$;
				\STATE $m_{t,2^{j+1}}^{(2 i)}=2 m_{t,2^{j}}^{(i)}$;
				\ENDFOR
				\ENDFOR
				\ENDFOR
				\ENDFOR
				\STATE Sort $m_{t,2^{n}}^{(i)}$ from~smallest to~largest;
				\STATE $\mathbf{u}_{\mathcal{F}}^{(p)}$ takes the~first $N-K$ values of~$m_{t,{2^{n}}}^{(i)}$
			\end{algorithmic}
		\end{algorithm}
		The MIMO polarization scheme of the previous subsection exhibited confidentiality limitations when the number of TAs is small. Consequently, we further explore potential methods of enhancing the system's confidentiality. As a benefit of the reciprocity of TDD systems, both parties have similar instantaneous gain values; however, the eavesdropper cannot obtain the legitimate link's instantaneous gain. Building on this concept, we model the gain $\mu_{\mathrm{t}}=\mathbf{h}_{\mathrm{t}}^{+} \mathbf{h}_{\mathrm{t}}$ of all RAs corresponding to the transmitter's $t$-th antenna and partition it into $P$ contiguous, but non-overlapping sub-intervals. In the Rayleigh fading channel model, the probability distribution function (PDF) of the gain $\mu$ for each TA can be expressed as:
		
		\begin{equation}
			\centering
			\label{PDF of gain}
			p(\mu)=\frac{1}{2^{T_{A}} \Gamma(T_{A})} x^{T_{A}-1} e^{-\mu / 2},
		\end{equation}
		where the Gamma function is $\Gamma(T_{A})=\int_{0}^{+\infty} \tau^{T_{A}-1} e^{-\tau} \mathrm{d} \tau$.
		
		Integrating the above equation yields $P$ continuous subintervals :
		
		\begin{equation}
			\centering
			\label{P of gain}
			\int_{\alpha_{p-1}}^{\alpha_{p}} \frac{1}{2^{T_{A}} \Gamma(T_{A})} \mu^{T_{A}-1} e^{-\mu / 2} d \mu=1 / P.
		\end{equation}
		
		Upon incorporating the channel gain segments into our MIMO polarization design, the different channel gain intervals map to distinct equivalent variances during the MIMO polarization process, subsequently yielding different coding methods, when matching the sub-channel reliability utilizing the classic GA algorithm, as outlined in Algorithm \ref{algorithm}. Moreover, the transmitter has $P$ unique coding methods for an identical detection order pattern. Table \ref{tab2} exemplifies the coding patterns for each sub-channel, when we have $P=16$ and a code length of $N=32$.
		
		\begin{table*}[!t]
			\centering
			\caption{Coding pattern for $P=16$ and $N=32$}
			\label{tab2}
			
			\setlength{\tabcolsep}{12mm}{
				\begin{tabular}{|c|c|c|}
					\hline
					{\color[HTML]{000000} \textbf{P}} & {\color[HTML]{000000} \textbf{Channel Gain Interval}}     & {\color[HTML]{000000} \textbf{Code Patterns}} \\ \hline
					{\color[HTML]{000000} 1}          & {\color[HTML]{000000} {[}0,1.4746)}                       & {\color[HTML]{000000} 1755}                   \\ \hline
					{\color[HTML]{000000} 2}          & {\color[HTML]{000000} {[}1.4746,1.8982)}                  & {\color[HTML]{000000} 5555}                   \\ \hline
					{\color[HTML]{000000} 3}          & {\color[HTML]{000000} {[}1.8982,2.2346)}                  & {\color[HTML]{000000} 5754}                   \\ \hline
					{\color[HTML]{000000} 4}          & {\color[HTML]{000000} {[}2.2346,2.5353)}                  & {\color[HTML]{000000} 115F}                   \\ \hline
					{\color[HTML]{000000} 5}          & {\color[HTML]{000000} {[}2.5353,2.8199)}                  & {\color[HTML]{000000} 017F}                   \\ \hline
					{\color[HTML]{000000} 6}          & {\color[HTML]{000000} {[}2.8199,3.0993)}                  & {\color[HTML]{000000} 1577}                   \\ \hline
					{\color[HTML]{000000} 7}          & {\color[HTML]{000000} {[}3.0993,3.3811)}                  & {\color[HTML]{000000} 107F}                   \\ \hline
					{\color[HTML]{000000} 8}          & {\color[HTML]{000000} {[}3.3811,3.6721)}                  & {\color[HTML]{000000} 5457}                   \\ \hline
					{\color[HTML]{000000} 9}          & {\color[HTML]{000000} {[}3.6721,3.9795)}                  & {\color[HTML]{000000} 1755}                   \\ \hline
					{\color[HTML]{000000} 10}         & {\color[HTML]{000000} {[}3.9795,4.3132)}                  & {\color[HTML]{000000} 3355}                   \\ \hline
					{\color[HTML]{000000} 11}         & {\color[HTML]{000000} {[}4.3132,4.6823)}                  & {\color[HTML]{000000} 1557}                   \\ \hline
					{\color[HTML]{000000} 12}         & {\color[HTML]{000000} {[}4.6823,5.1096)}                  & {\color[HTML]{000000} 5353}                   \\ \hline
					{\color[HTML]{000000} 13}         & {\color[HTML]{000000} {[}5.1096,5.6293)}                  & {\color[HTML]{000000} 70F1}                   \\ \hline
					{\color[HTML]{000000} 14}         & {\color[HTML]{000000} {[}5.6293,6.3184)}                  & {\color[HTML]{000000} FF00}                   \\ \hline
					{\color[HTML]{000000} 15}         & {\color[HTML]{000000} {[}6.3184,7.4166)}                  & {\color[HTML]{000000} 01F7}                   \\ \hline
					{\color[HTML]{000000} 16}         & {\color[HTML]{000000} {[}7.4166, +$\infty$)} & {\color[HTML]{000000}
						017F}                       \\ \hline
			\end{tabular}}
		\end{table*}
		
		The segmentation of channel gain not only compensates for the constraints of the MIMO polarization design scheme, but it even enhances the system's security. Under different detection sequences, distinct gain modes yield $T_{A}! \times P$ disparate coding schemes. However, the eavesdropper is unable to ascertain the detection sequence mode during the MIMO polarization process, nor can it obtain the legitimate link's instantaneous gain. Consequently, even if the eavesdropper acquires confidential information, it remain unaware of the correct coding structure, and thus, cannot achieve accurate decoding results.
		
		\section{Receiver Design}
		\label{Section Receiver Design}
		
		In this section, a detailed description of our receiver design employing MIMO polarization techniques is provided, along with an exposition of the processing steps for both the legitimate and eavesdropping parties.
		
		\subsection{Legitimate receiver}
		
		\begin{figure*}[!t]
			\centering
			\label{receiver}
			\includegraphics[width=0.8\textwidth]{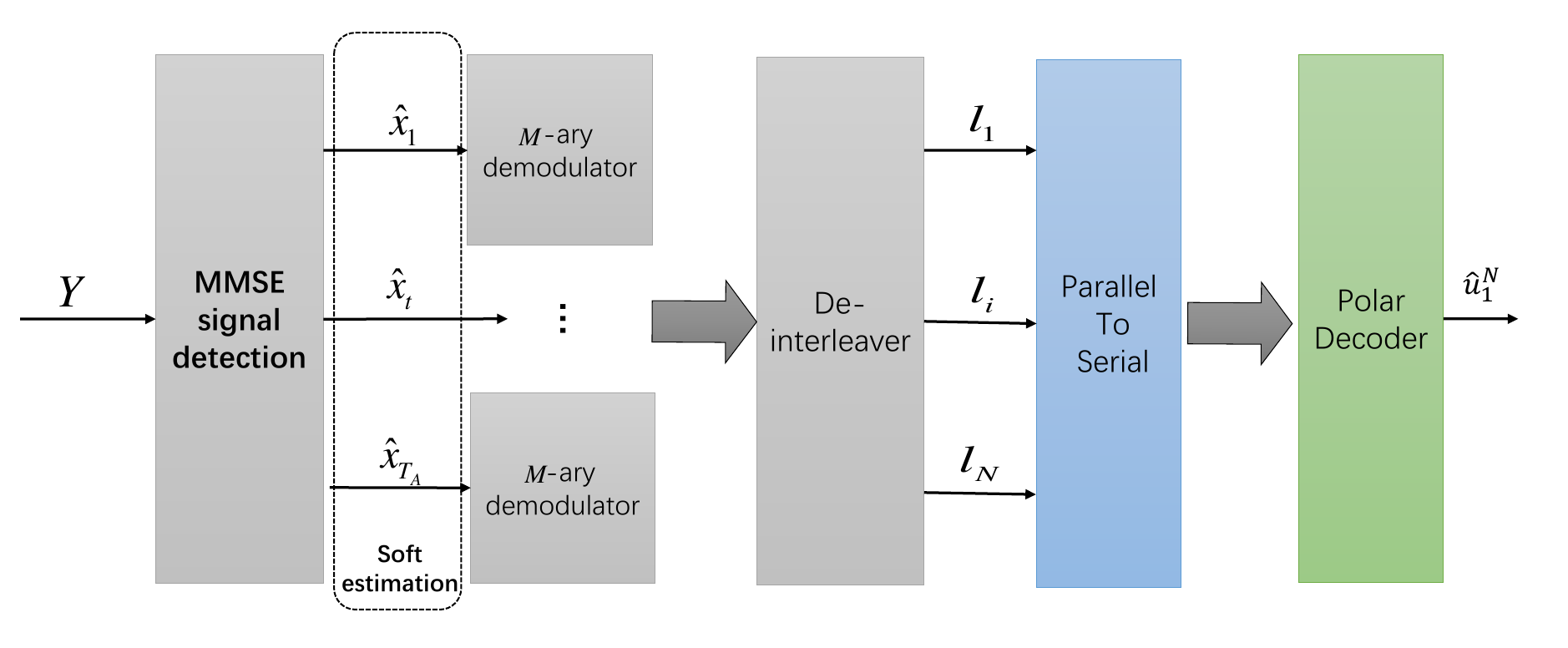}
			\caption{\label{Architecture based on MIMO polarisation design at the receiver.}Architecture based on our MIMO polarisation design at the receiver.}
		\end{figure*}
		
		
		
		For the legitimate user, a shared physical layer key exists for communication with the transmitter, enabling the acquisition of accurate MIMO detection sequence patterns and channel gain segmentation patterns. To minimize the processing latency and enhance the receiver performance attained, the legitimate receiver utilizes a minimum mean square error (MMSE) algorithm for concatenated MIMO detection and decoding. The MIMO detection's soft estimate is forwarded to the demodulator to derive the log-likelihood ratio (LLR), which is subsequently sent to the decoder for a hard decision, as illustrated in Fig. \ref{Architecture based on MIMO polarisation design at the receiver.}. The LLR expression is as follows \cite{PCandPLS1}:
		
		\begin{equation}
			\centering
			\label{llRs}
			\operatorname{LLR}_{\mathrm{B}}\left(b_{t,j}\right)=\ln \frac{\sum_{b_{t,j}=0} \exp \left(-\frac{\left\|\mathbf{y}_{\mathrm{B}}\left(b_{t,j}\right)-\left[\mathbf{h}_{1} \cdots \mathbf{h}_{N_{B}}\right] \mathbf{x}\left(b_{t,j}\right)\right\|_{\mathrm{F}}^{2}}{\sigma_{B}^{2}}\right)}{\sum_{b_{t,j}=1} \exp \left(-\frac{\left\|\mathbf{y}_{\mathrm{B}}\left(b_{t,j}\right)-\left[\mathbf{h}_{1} \cdots \mathbf{h}_{N_{B}}\right] \mathbf{x}\left(b_{t,j}\right)\right\|_{\mathrm{F}}^{2}}{\sigma_{B}^{2}}\right)},
		\end{equation}
		where $\mathbf{y}_{\mathrm{B}}(b_{t,j})$ represents the signal received by the legitimate receiver, while $\mathbf{x}(b_{t,j})$ denotes the modulation symbol comprising the transmitted bits $b_{t,j}$, and $\sigma_{B}^{2}$ is the noise variance of the legitimate link.
		
\begin{figure}[!t]
			\centering
			\includegraphics[width=0.49\textwidth]{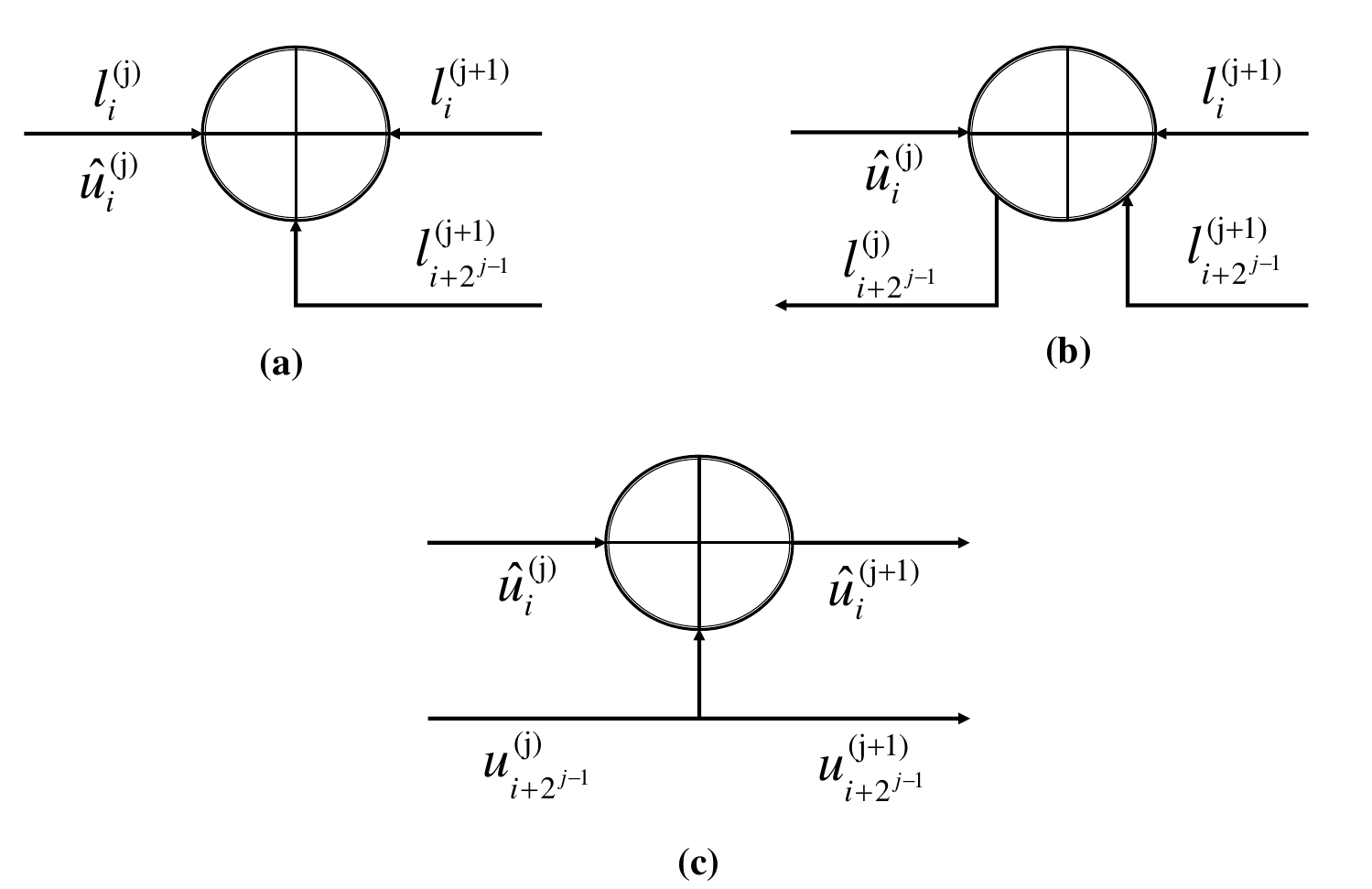}
			\caption{\label{SC decoder}. The SC decoding process for the mod-2 sum of the $i$-th and the $(i+2^{j-1})$-th bits at the $j$-th level: (a) the $f$ function, (b) the $g$ function and (c) partial sum calculation.}
		\end{figure}

		The LLRs are derived based on equation (\ref{llRs}) and subsequently they are input into the successive cancellation (SC) based stack polar decoder \cite{ref31} for making hard decisions, as depicted in Fig \ref{SC decoder}.
		
		Initially, the SC decoder carries out the operation seen in Fig \ref{SC decoder}(a), executing the $f$ function to the $(j+1)$-st layer using the $i$-th and $(i+2^{j-1})$-th LLRs on the left to obtain a new LLR, $l_{i}^{(j)}$. This can be expressed as:
		
		\begin{equation}
			\centering
			\label{f function}
			\begin{aligned}
				l_{i}^{(j)} &=f\left(l_{i}^{(j+1)}, l_{i+2 j^{j-1}}^{(j+1)}\right) \\
				&=2 \tanh ^{-1}\left(\tanh \left(l_{i}^{(j+1)} / 2\right) \tanh \left(l_{i+2^{j-1}}^{(j+1)} / 2\right)\right) \\
				& \approx \operatorname{sign}\left(l_{i}^{(j+1)}\right) \operatorname{sign}\left(l_{i+2^{j-1}}^{(j+1)}\right) \min \left(\left|l_{i}^{(j+1)}\right|,\left|l_{i+2^{j-1}}^{(j+1)}\right|\right)
			\end{aligned}
		\end{equation}
		The new LLR, $l_{i}^{(j)}$, is then subjected to hard decisions based on the coding structure of the legitimate link, which can be formulated as:
		
		\begin{equation}
			\centering
			\label{jude function}
			\hat{u}_{i}=\left\{\begin{array}{ll}
				0 & \text { if } l_{i}^{(1)} \geq 0 \text { or frozen bit } \\
				1 & \text { otherwise }
			\end{array}\right.
		\end{equation}
		
		Once the hard-decision based value of the $i$-th bit is determined, the LLRs $l_{i}^{(j+1)}$ and $l_{i+2^{j-1}}^{(j+1)}$ of the $(j+1)$-st layer are combined for executing the $g$ function, subsequently acquiring the soft information for the next bit. This is expressed as:
		
		\begin{equation}
			\centering
			\label{g function}
			l_{i+2^{j-1}}^{(j)}=\left\{\begin{array}{lc}
				l_{i+2^j}^{(j+1)}+x_{i}^{(j+1)} & \text { if } \hat{u}_{i}^{(j)}=0 \\
				l_{i+2^{(j-1)}}^{(j+1)}-x_{i}^{(j+1)} & \text { otherwise }.
			\end{array}\right.
		\end{equation}
		
		Likewise, the hard decision in Equation (\ref{jude function}) is executed based on the encoding structure of the legitimate link. Following this, $\hat{\mu}_{i}^{(j)}$ and $\hat{\mu}_{i+2^{j-1}}^{(j)}$ undergo XOR processing to derive $\hat{\mu}_{i}^{(j+1)}$, while $\hat{\mu}_{i+2^{j-1}}^{(j)}$ is directly transferred to $\hat{\mu}_{i+2^{j-1}}^{(j+1)}$. By iteratively performing the three operations depicted in Fig. \ref{SC decoder}, hard decisions are obtained for all transmitted bits, resulting in the final decoding outcome.
		
		Furthermore, to enhance the decoding capability of the legitimate link, the so-called successive cancellation list (SCL) and cyclic redundancy check (CRC)-SCL decoding algorithms of \cite{ref32} can be employed, which offer superior performance.
		
		As for the receiver design, the detector and decoder rely on a serial by concatenate construction. The computational overhead of the MMSE algorithm mainly depends on the dimension of the channel matrix and on the implementation of the algorithm, with a complexity order of $O[(T_{A}^2)]$ per symbol, where $T_{A}$ is the number of transmit antennas. Subsequently, the soft information representing the data is fed to the polarisation decoder, and the complexity of the SC decoder depends both on the number of iterations as well as on the dimensionality of the input data, which in our scheme has a complexity of $O[(\log(T_{A}))]$ per symbol. Specifically, the complexity per symbol in the proposed scheme may reach $O[(T_{A}^2 * \log(T_{A}))]$.
		
		The main reason for adopting the cascaded structure based on MMSE detection and SC decoding is that this receiver has both a low computational complexity as well as delay, which is favourable for employment in practical systems. In large-scale MIMO systems, this low-complexity and low-latency implementation is of pivotal significance.
		
		\subsection{Eavesdropper}
		
		As for the eavesdropper, an identical MMSE detection algorithm is employed for performing soft estimation of the intercepted signal. This is then entered into the demodulator to derive the soft LLRs, which can be expressed as:
		\begin{equation}
			\centering
			\label{llRs Eve}
			\operatorname{LLR}_{\mathrm{E}}\left(b_{t,j}\right)=\ln \frac{\sum_{b_{t,j}=0} \exp \left(-\frac{\left\|\mathbf{y}_{\mathrm{E}}\left(b_{t,j}\right)-\left[\mathbf{g}_{1} \cdots \mathbf{g}_{N_{E}}\right] \mathbf{x}\left(b_{t,j}\right)\right\|_{\mathrm{F}}^{2}}{\sigma_{E}^{2}}\right)}{\sum_{b_{t,j}=1} \exp \left(-\frac{\left\|\mathbf{y}_{\mathrm{E}}\left(b_{t,j}\right)-\left[\mathbf{g}_{1} \cdots \mathbf{g}_{N_{E}}\right] \mathbf{x}\left(b_{t,j}\right)\right\|_{\mathrm{F}}^{2}}{\sigma_{E}^{2}}\right)},
		\end{equation}
		where $\mathbf{y}_{\mathrm{E}}(b_{t,j})$ represents the signal received by the eavesdropper, Eve, while $\mathbf{x}(b_{t,j})$ represents the modulation symbol comprising the transmitted bits $b_{t,j}$ and $\sigma_{E}^{2}$ is the noise variance of Eve's link.
		
		Subsequently, these LLRs are fed into the decoder for error correction. On one hand, Eve is incapable of obtaining the antenna detection sequence pattern during the MIMO polarization of the legitimate link. She only has a $1/T{A}!$ probability of acquiring the correct detection pattern, which prevents her from inferring the variance of the equivalent fading channel or the coding structure of the legitimate link. On the other hand, even when the transmitter has a limited number of antennas, the eavesdropper is unable to determine the channel gain range of the legitimate link, which also prevents her from acquiring the coding structure of the legitimate link. The PLS framework, based on our MIMO polarization design combined with the channel gain segmentation based design, enhances the performance of the legitimate link, while significantly degrading the eavesdropper's success probability.
		
		\section{Secrecy rate analysis}
		\label{Section Secrecy rate analysis}
		
		In this section, the secrecy rate for the proposed scheme is analyzed under both Gaussian-distributed input and finite-alphabet input scenarios. The secrecy rate is defined as the positive difference between the maximum achievable data rates of the legitimate and eavesdropping links.
		
		\subsection{Gaussian-distributed input}
		Under the Gaussian-distributed input condition, it is assumed that the signal transmitted by the legitimate link obeys the complex Gaussian distribution $\mathcal{C N}\left(0, \sigma_{B}^{2}\right)$. Based on the above secrecy rate definition, the secrecy rate under the Gaussian-distributed input condition is formulated as:
		\begin{equation}
			\centering
			I_{PLS}=\max \left\{0, I\left(\mathbf{W}_{B}\right)-I\left(\mathbf{W}_{E}\right)\right\},
		\end{equation}
		where $I(\mathbf{W}_{B})$ and $I(\mathbf{W}_{E})$ denote the channel capacities of the legitimate and eavesdropping links, respectively.
		
		Since the instantaneous gain of the channel is discretised, the channel capacities of the legitimate and eavesdropping links under Gaussian-distributed input conditions can be further expressed as:
		
		\begin{equation}
			\centering
			\label{I_W_B}
			I\left(\mathbf{W}_{B}\right)=\frac{1}{P} \cdot\sum_{p=1}^{P} I\left(\mathbf{W}_{B}\right)^{(p)},
		\end{equation}
		\begin{equation}
			\centering
			I\left(\mathbf{W}_{E}\right)=\frac{1}{P} \cdot \sum_{p=1}^{P} I\left(\mathbf{W}_{E}\right)^{(p)},
		\end{equation}
		where $P$ represents the number of gain segments. Furthermore,  $I(\mathbf{W}_{B})^{(p)}$ and $I(\mathbf{W}_{E})^{(p)}$ correspond to the channel capacities of the legitimate and eavesdropping links, when the channel gain falls within the $p$-th interval.
		
		Furthermore, for a specific channel gain interval, following the transmitter's MIMO, modulation and bit polarization, the symmetric capacity expression becomes:
		
		\begin{equation}
			\centering
			I\left(\mathbf{W}_{B}\right)^{(p)}=S \cdot \sum_{t=1}^{T_{A}} I\left(\mathbf{W}_{t}\right)^{(p)}=S \cdot \sum_{t=1}^{T_{A}} \sum_{j=1}^{m} I\left(\mathbf{W}_{t, j}\right)^{(p)},
		\end{equation}
		where $S$ represents the total number of transmission time slots and $m$ denotes the number of bits contained in each modulation symbol. Furthermore, $I(\mathbf{W}_{t, j})^{(p)}$ is the capacity of the MIMO-polarised bit sub-channel, which is given by:
		
		\begin{equation}
			\centering
\begin{aligned}
			&I\left(\mathbf{W}_{t, j}\right)^{(p)} \\
&=\sum_{b_{t,j}} \int_{-\infty}^{+\infty} \int_{-\infty}^{+\infty} \frac{1}{2^{j}} p_{t}\left(y_{B} \mid b_{t,j}\right) \cdot \log \frac{p_{t}\left(y_{B} \mid b_{t,j}\right)}{p_{t}\left(y_{B} \mid 1 \right) p_{t}\left(y_{B} \mid 0 \right)}  d u d v,
\end{aligned}
		\end{equation}
		where $y_{B}$ denotes the received signal, and $u=\mathfrak{R}\left(\mathrm{y}_{B}\right),  v=\mathfrak{S}\left(\mathrm{y}_{B}\right)$. Furthermore, under the Gaussian-distributed input condition, the expression for $p_{t}\left(y_{B} \mid b_{t,j}\right)$ is:
		
		\begin{equation}
			\centering
			p_{t}\left(y \mid b_{t,j} \right)=\frac{1}{2^{m-j}} \sum_{x_{t}} \frac{1}{\pi \sigma_{B}^{2}} \cdot \exp \left(\frac{-\left\|y_{B}-x_{t}\right\|^{2}}{\sigma_{B}^{2}}\right),
		\end{equation}
		where $x_{t}$ denotes the $t$-th antenna's transmitted signal in the legitimate link.
		
		Simultaneously, the eavesdropper is unaware of the transmitter's specific MIMO-polarization design process, implying that it will encounter $T_{A}!$ signal detection patterns. Thus, the eavesdropper has a maximum probability of inferring the correct pattern given by $1/T_{A}$!, Hence the channel capacity of the eavesdropping link becomes:
		
		\begin{equation}
			\centering
			I\left(\mathbf{W}_{E}\right)^{(p)}=\frac{S}{T_{A}!} \cdot  \sum_{t=1}^{T_{A}} I\left(\mathbf{W}_{t}\right)^{(p)}.
		\end{equation}
		
		Consequently, under the Gaussian-distributed input condition, the system's secrecy rate can be reformulated as:
		
		\begin{equation}
			\centering
			I_{PLS}=\max \left\{0, I\left(\mathbf{W}_{B}\right)-\frac{1}{T_{A} !} \cdot I\left(\mathbf{W}_{B}\right)\right\},
		\end{equation}
		where $I\left(\mathbf{W}_{B}\right)$ is provided by Equation (\ref{I_W_B}).
		
		\subsection{Finite-Alphabet Input}
		Taking into account a more practical scenario, the secrecy rate is formulated under finite symbol input conditions, representing the maximum positive difference between the achievable rates of the legitimate and eavesdropping links. To consolidate the expressions, we assume that the transmitter's transmit power is $\hat{\sigma}_{B}^{2}$, resulting in the secrecy rate expression:
		
		\begin{equation}
			\centering
			R_{\mathrm{PLS}}=\max \left(0, R_{B}-R_{E}\right),
		\end{equation}
		where $R_{B}$ denotes the legitimate link's maximum achievable rate, while $R_{E}$ represents the eavesdropper's  maximum achievable rate.
		
		As the transmit power increases, an upper bound on the legitimate link's achievable rate can be formulated as:
		\begin{equation}
			\centering
			\label{R_B_upbound}
			\lim _{\mathrm{\hat{\sigma}_{B}}^{2} \rightarrow+\infty} {R}_{\mathrm{B}}=T_{A} \cdot \log _{2} M.
		\end{equation}
		
		Based on equation (\ref{R_B_upbound}), for simplicity, we disregard the time index and express the legitimate link's achievable rate \cite{SM-LIU} under a given channel as:
		
		\begin{equation}
			\centering
            \begin{aligned}
			&R_{B}=T_{A} \cdot  \log _{2} M-\frac{1}{T_{A} \cdot M} \sum_{t=1}^{T_{A}} \sum_{k=1}^{M} E \\\\
& \left\{\log _{2} 
 \left(1+\sum_{\substack{t=1 \\ t \neq t^{\prime}}}^{T_{A}} \exp \left(-\rho\left[\left(\mathbf{v}_{\mathrm{t}, \mathrm{t^{\prime}}}+\mathbf{z}_{B}\right)^{\dagger}\left(\mathbf{v}_{\mathrm{t} ,\mathrm{t^{\prime}}}+\mathbf{z}_{B}\right)-\mathbf{z}_{B}^{\dagger} \mathbf{z}_{B}\right]\right)\right\}\right\},
            \end{aligned}
		\end{equation}
		where $\mathbf{v}_{t,t^{\prime}}=\mathbf{H}_{t} x_{k}-\mathbf{H}_{t^{\prime}} x_{k} $, $\mathbf{H}_{t}$ represents the first column through the $t$-th column of the original MIMO matrix $\mathbf{H}$ and $\rho=\hat{\sigma}_{B}^{2}/\sigma_{B}^{2}$ denotes the SNR.
		
		Similarly, for the eavesdropper, there is only a $T_{A}!$ probability of inferring the correct MIMO detection sequence pattern. Hence, the eavesdropping link's achievable rate under this condition is expressed as:
		
		\begin{equation}
			\centering
			\label{R_E}
			R_{E}=\frac{1}{T_{A}!} \cdot R_{B}.
		\end{equation}
		
		Thus, under the finite-alphabet input condition, the system's secrecy rate can be reformulated as:
		
		\begin{equation}
			\centering
			R_{PLS}=\max \left\{0, R_{B}-\frac{1}{T_{A} !} \cdot R_{B}\right\}
		\end{equation}
		
		As demonstrated by the aforementioned equation, as the number of transmit antenna and the power increase, the system's secrecy rate approaches the legitimate link's achievable rate. The eavesdropper's achievable rate is substantially reduced, resulting in a relatively high secrecy rate for the system.
		
		\section{Simulation result}
		\label{Section Simulation result}
		In this section, we initially confirm that the proposed scheme exhibits a substantial performance enhancement compared to the conventional MIMO system. Then, we compare the performance of authorized users and eavesdroppers both in terms of their BER and BLER, thereby establishing the scheme's security enhancement. Subsequently, we present numerical results for the secrecy rate of the proposed method, considering both Gaussian distributed and discrete symbol input, which substantiates the efficiency of this approach. The simulation parameters are shown in Table \ref{tab3 simulation parameters}.
		
				\begin{table}[]
			\centering
			\caption{Simulation parameters}
			\label{tab3 simulation parameters}
			\setlength{\tabcolsep}{0.2mm}{
				\begin{tabular}{|l|r|}
					\hline
					{ \textbf{Parameters}}                           & {\textbf{Values}}                    \\ \hline
					{ Number of transmitter antennas $T_{A}$}                & { 2,4,8}                              \\ \hline
					{ Number of receiver antennas for legitimate $N_{B}$}   & { 1,2,4,8}                            \\ \hline
					{ Number of receiver antennas for eavesdropper $N_{E}$} & {1,2,4,8}                            \\ \hline
					{ Length of polar code $N$}                                & { 512,1024}                           \\ \hline
					{ Length of information bits $K$}                    & {256,512}                            \\ \hline
					{ Number of channel segments $P$}                    & {1,4,8,16,32}                           \\ \hline
					{ MQAM modulation order $M$}                         & {2,4,16}                             \\ \hline
					{ Number of elements in the lists $L$}                               & {16}                                 \\ \hline
					{ Number of CRC bits}                            & {24}                                 \\ \hline
					{ Channel model}                                 & {Rayleigh} \\ \hline
			\end{tabular}}
		\end{table}
		
		\subsection{BER and BLER performance}
		
			\begin{figure}[tbp]
			\centering
			\includegraphics[width=0.49\textwidth]{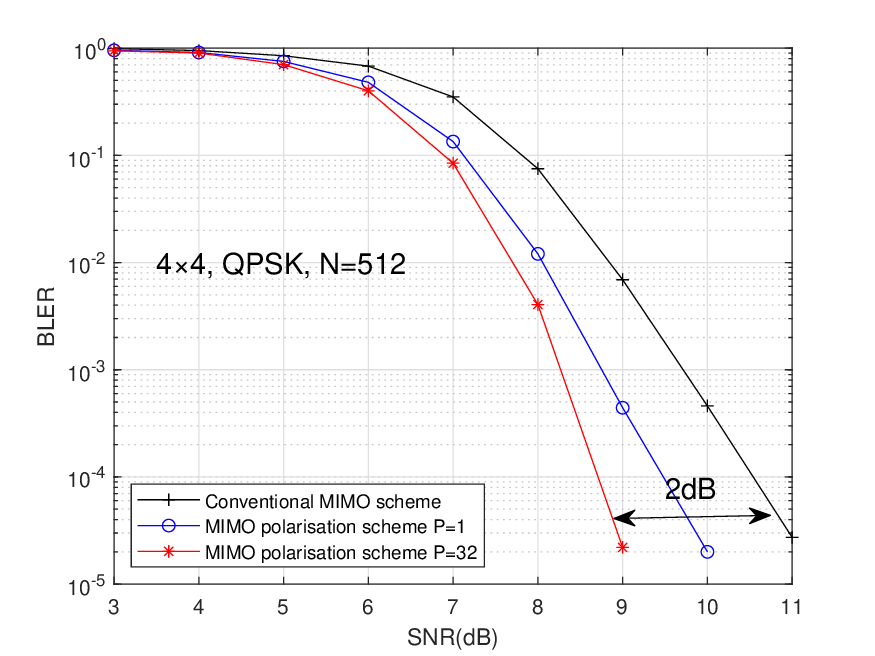}
			\caption{\label{figure/MIMO-polarisation-scheme}BLER performance based on MIMO-polarization system versus conventional MIMO system.}
		\end{figure}
		
		\begin{figure}[!htb]
			\centering
			\subfigure[BER performance under $M=4$.\label{QPSK-512-BER-4X4 a}]
			{\includegraphics[width=0.5\textwidth]{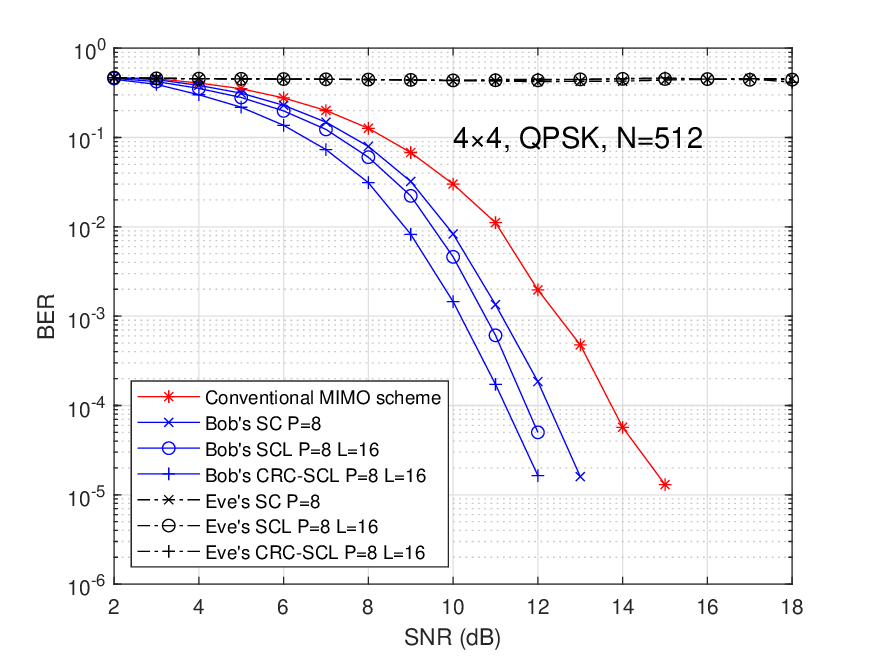}}
			\subfigure[BER performance under $M=16$. \label{QPSK-512-BER-4X4 b}]
			{\includegraphics[width=0.5\textwidth]{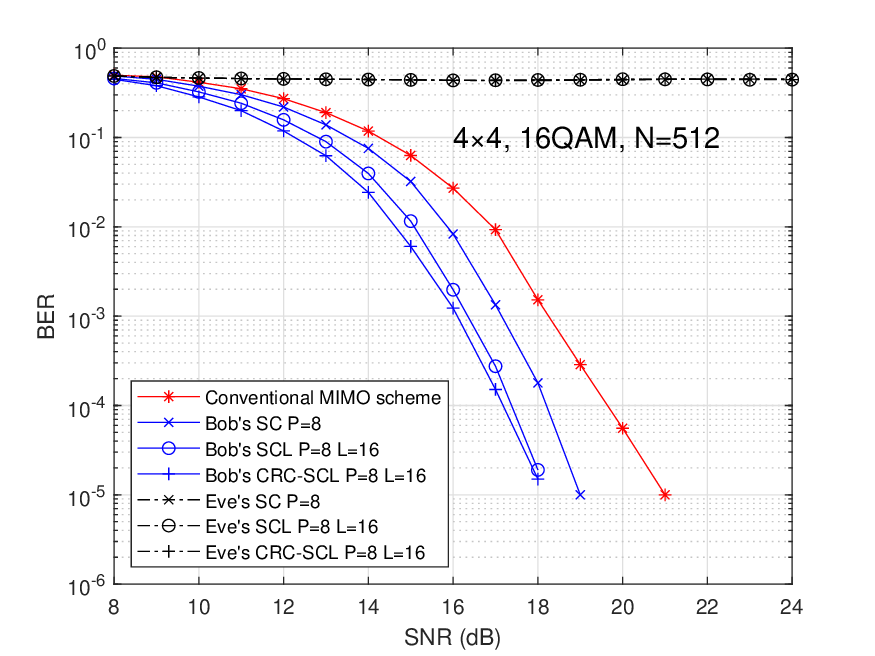}}
			
			\caption{BER performance at Bob and Eve, where $N=512, P=8$ and $T_{A}=N_{B}=N_{E}=4$. (a) QPSK, (b) 16QAM.}
			\label{BER N=512}
		\end{figure}
		
		As depicted in Fig. \ref{figure/MIMO-polarisation-scheme}, MIMO-polarization transmission, modulation-polarization and bit-level polarization scheme, yields substantial performance improvements compared to conventional MIMO transmission. Explicitly, when we set the number of instantaneous channel gain intervals to $P=32$, our scheme provides an improvement of about $2$dB over a conventional MIMO scheme at BLER $\approx 2 \cdot10^{-5}$ . This enhancement is attributed to the increased polarization effect attained by our multi-domain polarization system, leading to improved bit sub-channel reliability and more secure confidential information transmission for a given code length.
		
		Fig. \ref{QPSK-512-BER-4X4 a} characterizes the BER of both the legitimate party and of the eavesdropper, given a code length of $N=512$. The number of instantaneous channel gain intervals was set to $P=8$, and $4$ transmit and receive antennas were used. Fig. \ref{QPSK-512-BER-4X4 a} employs QPSK modulation, illustrating that as the SNR increases, Bob's BER is reduced rapidly, while Eve's BER remains approximately 0.5. When the high-performance decoding algorithms are employed \citen{ref32}, the legitimate party's BER improves, further, but the eavesdropper fails to glean any useful information. Comparable results are observed also for 16QAM, as shown in Fig. \ref{QPSK-512-BER-4X4 b}, which validates the benefits of the proposed scheme. Furthermore, it can be observed in Fig. \ref{BER N=512} that the performance of the legitimate link is improved compared to the conventional MIMO scheme.
		
\begin{figure}[!htb]
			\centering
			\subfigure[BER performance under $M=4$.\label{QPSK_1024_BER_8X8 a}]
			{\includegraphics[width=0.5\textwidth]{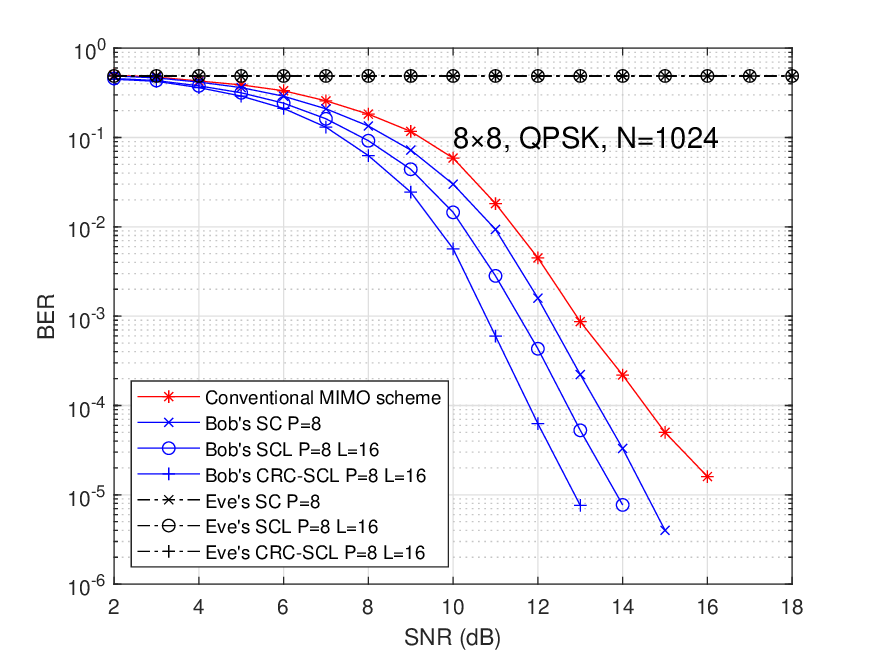}}
			\subfigure[BER performance under $M=16$. \label{16QAM_1024_BER_8X8 b}]
			{\includegraphics[width=0.5\textwidth]{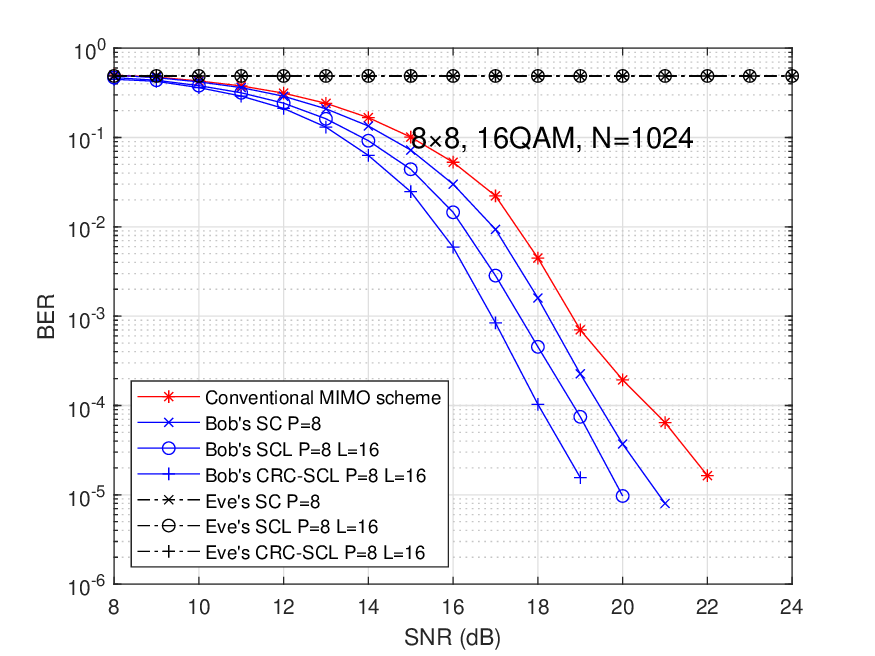}}
			
			\caption{BER performance at Bob and Eve, where $N=1024, P=8$ and $T_{A}=N_{B}=N_{E}=8$. (a) QPSK, (b) 16QAM.}
			\label{BER N=1024}
		\end{figure}

		
		Let us how explore the impact of increasing the number of antennas and the code length, while enhancing the code length is known to improve the error correction performance of polar codes. Fig. \ref{BER N=1024} demonstrates the decoding performance when the code length is $N=1024$, the number of instantaneous channel gain intervals is $P=8$, and the number of transmit and receive antennas is $8$. The trend observed aligns with that of Fig. \ref{BER N=512}. Regardless of whether high-order or low-order modulation is employed, the eavesdropper's bit error rate remains approximately 0.5, showing no improvement.Upon increasing the SNR, this is a testimony to the reliability of our PLS scheme based on MIMO-polarization.
		
		\begin{figure}[tbp]
			\centering
			\includegraphics[width=0.5\textwidth]{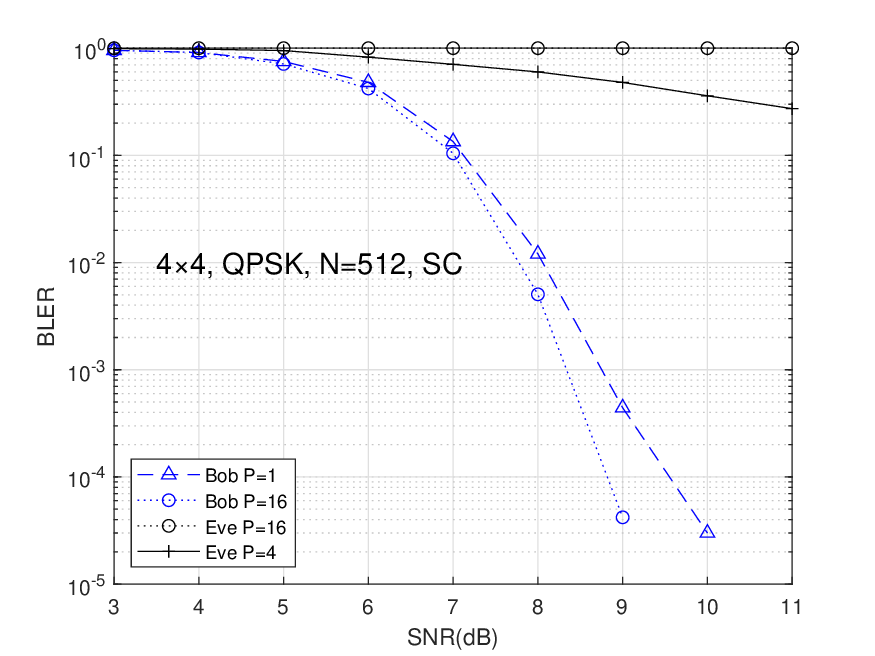}
			\caption{\label{performance of Bob and Eve at different P values}BLER performance of Bob and Eve for different $P$ values}
		\end{figure}
	
	\begin{figure}[tbp]
		\centering
		\includegraphics[width=0.5\textwidth]{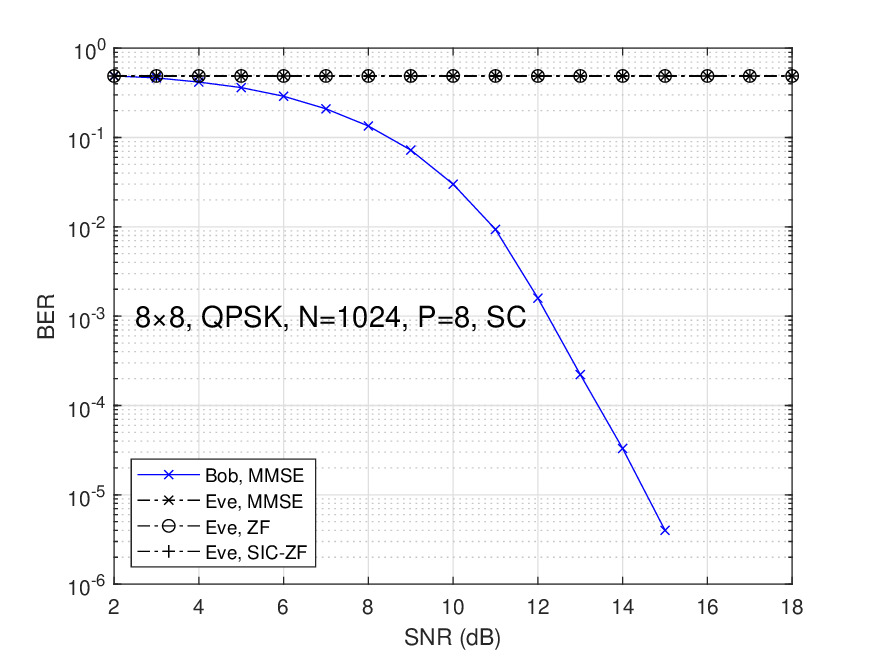}
		\caption{\label{BER-different-detection-algorithm}BER performance at Bob and Eve, where Eve used different detection algorithms and $N=1024, P=8$ and $T_{A}=N_{B}=N_{E}=8$.}
	\end{figure}
		
		 Fig. \ref{performance of Bob and Eve at different P values} examines the influence of the number of channel gain intervals on the BLER of both the legitimate and eavesdropping links. As the number of intervals increases, the legitimate link's BLER performance improves, while the eavesdropper's performance degrades. Exploiting the segmented channel gain enhances the key randomness, making it more challenging for the eavesdropper to infer any useful information.
		
		 In order to characterize the achievable security performance of this scheme, we added simulation results, where the eavesdropper uses different detection algorithms. As shown in Fig.\ref{BER-different-detection-algorithm}, the eavesdropper still fails to decode a complete frame when using the zero forcing (ZF) detection algorithm and the serial interference cancellation (SIC-ZF) detection algorithm.
		
		 By observing Figs. \ref{figure/MIMO-polarisation-scheme}, \ref{BER N=512}, \ref{BER N=1024}, \ref{performance of Bob and Eve at different P values}, \ref{BER-different-detection-algorithm}, it becomes evident that the PLS scheme based on MIMO-polarization attains significant performance improvement, compared to conventional MIMO transmission.
		
		\subsection{Secrecy-rate results}
		In this subsection, we characterize the secrecy rate of the proposed scheme, with $I_{B}$ denoting the channel capacity of the legitimate link, and $I_{P}$ representing the system's secrecy rate.
		\subsubsection{Gaussian distributed input}
		
		\begin{figure}[tbp]
			\centering
			\includegraphics[width=0.5\textwidth]{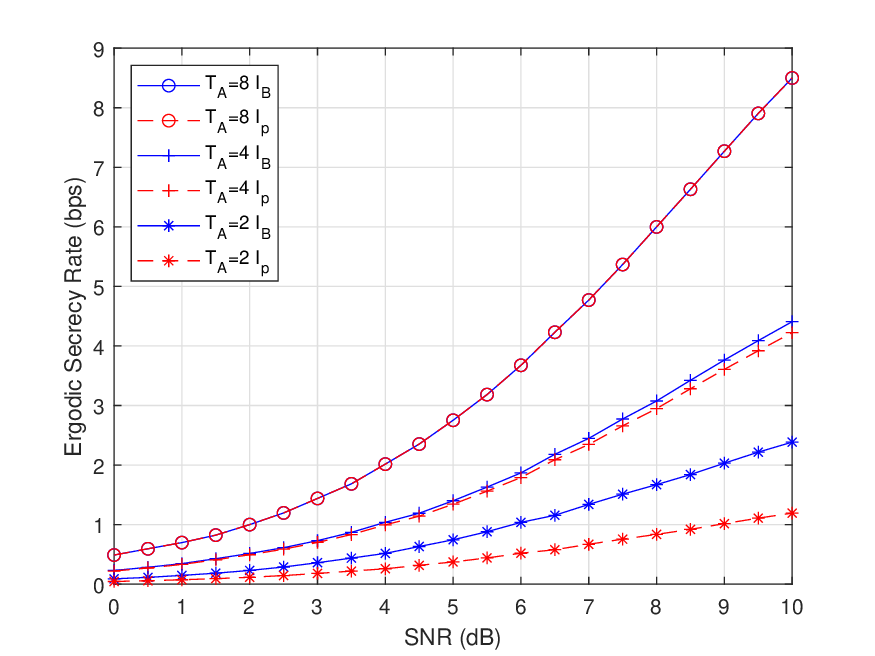}
			\caption{\label{Secrecy-rate-capacity-one}The ergodic secrecy rate for Gaussian-distributed input, where $N=1024, P=8$, $N_{B}=N_{E}=T_{A}$ and QPSK is used. }
		\end{figure}
\begin{figure}[tbp]
			\centering
			\includegraphics[width=0.5\textwidth]{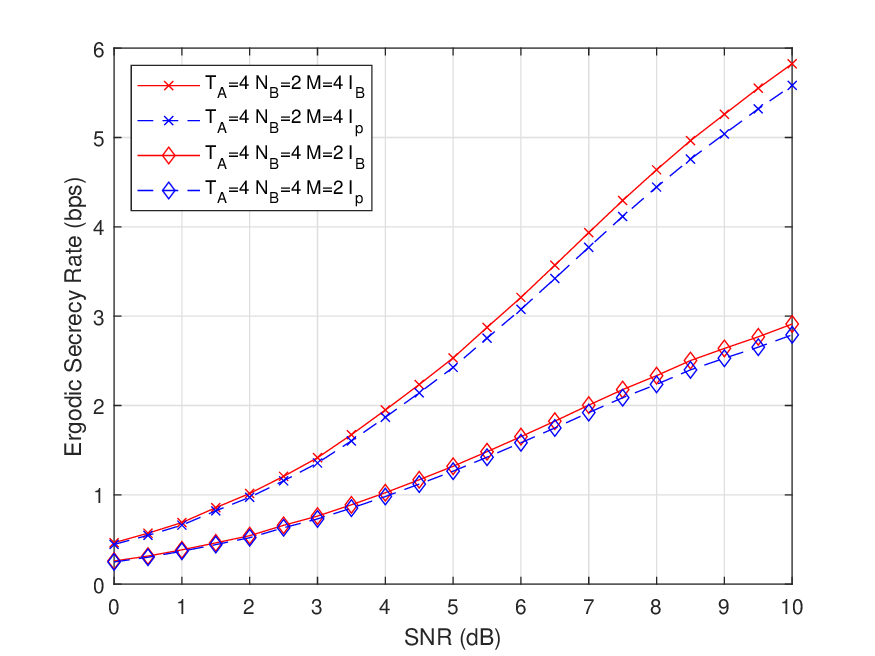}
			\caption{\label{Secrecy-rate-capacity-two}The ergodic secrecy rate for Gaussian-distributed input, where $N=1024, P=8$ and $N_{B}$=$N_{E}$.}
		\end{figure}

		Under the Gaussian distribution input condition, as depicted in Fig. \ref{Secrecy-rate-capacity-one}, the secrecy rate of the proposed scheme approaches the channel capacity of the legitimate link, as the number of transmit antennas increases. Notably, when $T_{A} =8$, the two values essentially coincide, demonstrating that the eavesdropper's decoding performance is significantly degraded under these conditions, ensuring the system's confidentiality. Additionally, the influence of the number of receive antennas and of the modulation scheme is also investigated. As illustrated in Fig. \ref{Secrecy-rate-capacity-two}, the system's secrecy rate using BPSK is lower than that of QPSK, which is consistent with our theoretical expectations. Under both modulation schemes, the system's secrecy rate is very close to the legitimate link's channel capacity, confirming the system's practicality. Upon scrutinising Fig. \ref{Secrecy-rate-capacity-one} and Fig. \ref{Secrecy-rate-capacity-two}, it becomes apparent that increasing the number of receive antennas, given the same number of transmit antennas and modulation scheme, has a certain impact on the system's rate due to the prior influence of data flow  and interference from other antennas, which aligns with the theory.
		
		Overall, under the Gaussian distributed input condition, the system's secrecy rate approaches the channel capacity of the legitimate link, as the number of transmit antennas increases, regardless of the choice of modulation scheme or the number of receive antennas. This observation is in line with the previously discussed BER performance and further validates the reliability of the proposed scheme.
		
		\subsubsection{Finite-Alphabet Input}
		
		\begin{figure}[tbp]
			\centering
			\includegraphics[width=0.5\textwidth]{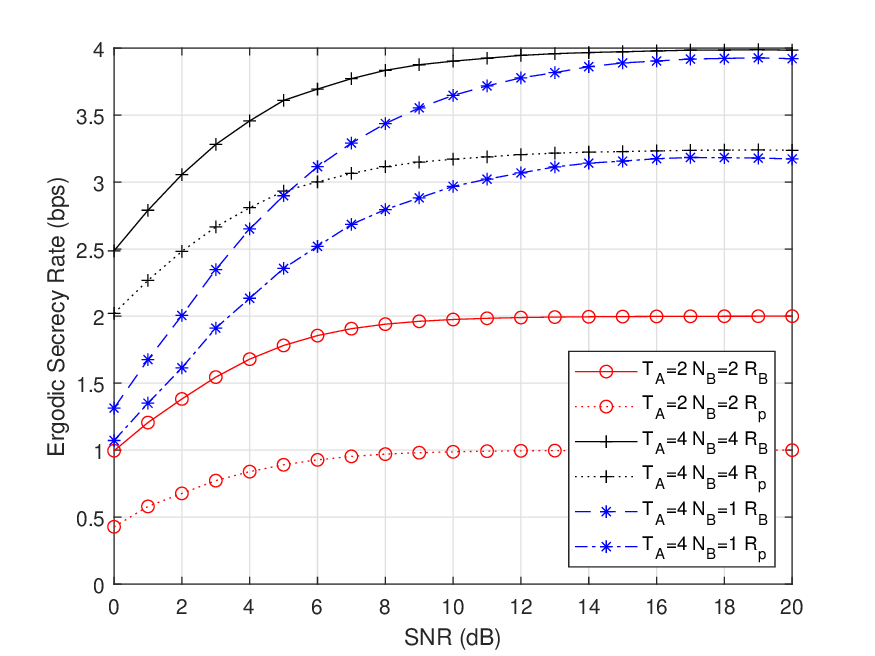}
			\caption{\label{Secrecy-rate-sulv-one}The ergodic secrecy rate for Finite-Alphabet Input, where $N=1024, P=8$ and BPSK is used.}
		\end{figure}
		
		In a more practical scenario, under the finite-alphabet input condition, this section presents the maximum achievable rate for both the legitimate link and the system. As depicted in Fig. \ref{Secrecy-rate-sulv-one}, $R_{B}$ represents the legitimate link's achievable rate, and $R_{p}$ denotes the system's confidential achievable rate. Upon increasing the number of transmit antennas, the system's achievable rate gradually approaches that of the legitimate link, exhibiting a similar trend to that observed under the Gaussian distributed input condition, which substantiates the scheme's reliability. Furthermore, for the same number of transmit antennas, reducing the number of receive antennas has some impact on the system rate, but as the SNR increases, both upper limits become identical. This consistency with the theory does not affect the difference between the secrecy rate and the legitimate link's achievable rate.
		
		\begin{figure}[tbp]
			\centering
			\includegraphics[width=0.5\textwidth]{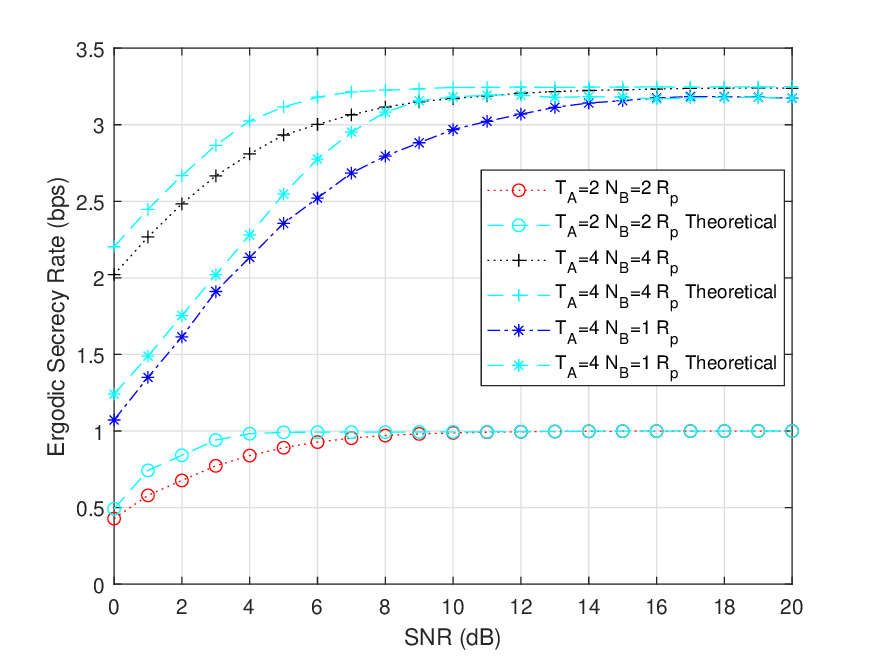}
			\caption{\label{Secrecy-rate-sulv-two}The ergodic secrecy rate for Finite-Alphabet Input, also showing theoretical values related to BER, where $N=1024, P=8$ and BPSK is used.}
		\end{figure}
		
\begin{figure}[tbp]
	\centering
	\includegraphics[width=0.5\textwidth]{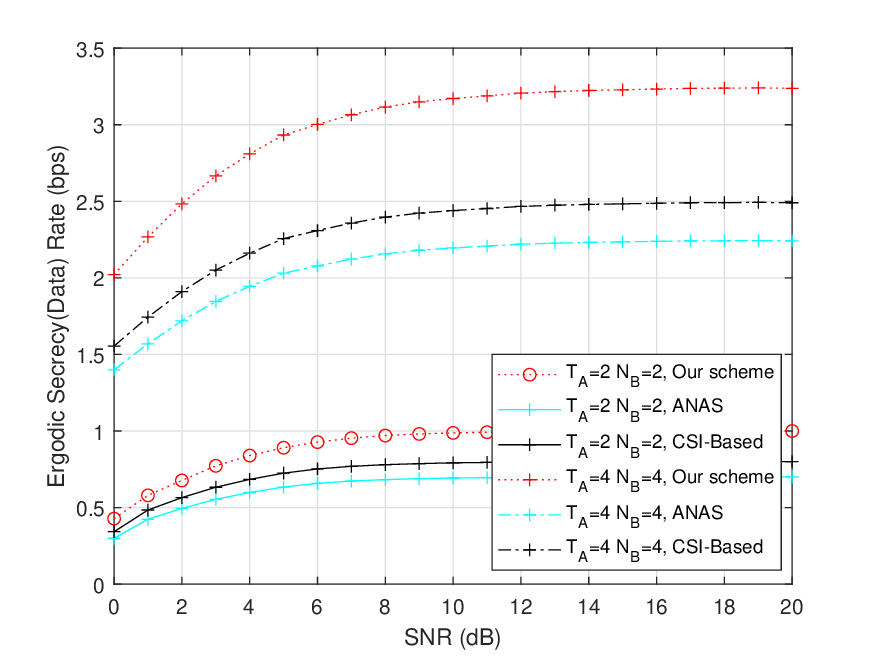}
	\caption{\label{Secrecy-rate-sulv-VS-AN-CSI}The ergodic secrecy rate for Finite-Alphabet Input, also showing comparisons with existing schemes , where $N=1024, P=8$ and BPSK is used.}
\end{figure}
		
		Additionally, to verify that our multi-domain polarization-based design can enhance the system's overall polarization effect, the BER performance and secrecy rate are jointly analyzed. Under the same conditions, the legitimate link and eavesdropper's BER values are substituted into the binary symmetric channel (BSC) to obtain the secrecy rate as the theoretical value in the current situation. This is because polar codes have been shown to achieve the theoretical channel capacity of BSC. Upon comparing this theoretical value to the system's secrecy rate, wo can see in Fig. \ref{Secrecy-rate-sulv-two}, that the difference between the two secrecy rates is minimal, and they converging as the SNR increases. This result demonstrates that the proposed PLS scheme based on our multi-domain polarization design approaches the theoretical value under Rayleigh channel conditions, further corroborating the advantages of this approach.
		
		To further validate the potential of the proposed scheme, We added a comparison to the above two schemes \cite{AN-Based-VS,CSI-Based-VS}, as shown in Fig. \ref{Secrecy-rate-sulv-VS-AN-CSI}, observe that the ergodic secrecy rate of our proposed scheme is higher than that of the above two schemes. Compared to the AN and CSI based schemes, our scheme improves the secrecy rate of the system despite its reduced a overhead, which verifies the effectiveness of the proposed scheme.

		\section{Conclusions}
		\label{Section Conclusion}
		
		 A novel physical layer security framework was conceived by leveraging both MIMO, modulation, and bit polarization. The proposed framework improves the legitimate link's performance, while significantly degrading the eavesdropper's reception to the point, where correctly decoding a complete data frame becomes nearly impossible. Furthermore, the channel's instantaneous gain is partitioned into segments to increase the key's randomness, hence again, improving the legitimate link's performance and degrading the eavesdropper's reception capability. The scheme's reliability is validated through simulations. Moreover, the system's secrecy rate is examined, and the numerical results demonstrate the scheme's confidentiality. It is worth mentioning that the receiver uses a simple cascaded design, and we will consider proposing more complex receiver architectures with better performance in our future work.
		
		\bibliography{reference/reference}
	\end{document}